\documentclass[useAMS,usenatbib]{mn2e}
\usepackage{graphicx}

\newcommand{\Teff}{\mbox{$T_{\mathrm{eff}}$}}
\newcommand{\SDSS}{SDSS\,J121258.25-012310.1}
\newcommand{\sdss}{SDSS\,J1212-0123}
\newcommand{\gkvir}{GK\,Vir}
\newcommand{\heii}{He\,{\sc ii}}
\newcommand{\caii}{Ca\,{\sc ii}}
\newcommand{\mgii}{Mg\,{\sc ii}}
\newcommand{\kms}{\mbox{km}\,\mbox{s}^{-1}}
\newcommand{\msy}{\mbox{$\mathrm{M_{\sun}\,yr^{-1}}$}}

\title[Precise parameters for SDSS J1212-0123 and GK Vir]{A precision study of two eclipsing white dwarf plus M dwarf binaries}

\author[S. G. Parsons et al.]{S.~G.~Parsons$^{1}$\thanks{steven.parsons@warwick.ac.uk},
T.~R.~Marsh$^{1}$,
B.~T.~G{\"a}nsicke$^{1}$,
A.~Rebassa-Mansergas$^{2}$,
\newauthor
V.~S.~Dhillon$^{3}$,
S.~P.~Littlefair$^{3}$,
C.~M.~Copperwheat$^{1}$,
R.~D.~G.~Hickman$^{1}$,
\newauthor
M.~R.~Burleigh$^{4}$,
P.~Kerry$^{3}$,
D.~Koester$^{5}$,
A.~Nebot G{\'o}mez-Mor{\'a}n$^{6}$,
S.~Pyrzas$^{1}$,
\newauthor
C.~D.~J.~Savoury$^{3}$,
M.~R.~Schreiber$^{2}$,
L.~Schmidtobreick$^{7}$,
A.~D.~Schwope$^{6}$,
\newauthor
P.~R.~Steele$^{4}$
and C.~Tappert$^{2}$\\
$^{1}$Department of Physics, University of Warwick, Coventry, CV4 7AL, UK \\
$^{2}$Departmento de F{\'i}sica y Astronom{\'i}a, Universidad de Valpara{\'i}so, Avenida Gran Bretana 1111, Valpara{\'i}so, Chile \\
$^{3}$Department of Physics and Astronomy, University of Sheffield, Sheffield S3 7RH, UK\\
$^{4}$Department of Physics and Astronomy, University of Leicester, Leicester, LE1 7RH, UK\\
$^{5}$Institut f{\"u}r Theoretische Physik und Astrophysik, Universit{\"a}t Kiel, Germany\\
$^{6}$Leibniz-Institut f{\"u}r Astrophysik Potsdam, An der Sternwarte 16, 14482 Potsdam, Germany\\
$^{7}$European Southern Observatory, Alonso de Cordova 3107, Santiago, Chile}
\begin{document}
\input{pjw_aas_macros.cls}
\date{Accepted 2011 November 23. Received 2011 November 23; in original form 2011 October 3}

\pagerange{\pageref{firstpage}--\pageref{lastpage}} \pubyear{2011}

\maketitle

\label{firstpage}

\begin{abstract}

We use a combination of X-shooter spectroscopy, ULTRACAM high-speed photometry and SOFI near-infrared photometry to measure the masses and radii of both components of the eclipsing post common envelope binaries {\SDSS} and {\gkvir}. For both systems we measure the gravitational redshift of the white dwarf and combine it with light curve model fits to determine the inclinations, masses and radii. For {\sdss} we find an inclination of $i=85.7^{\circ}\pm0.5^{\circ}$, masses of $M_\mathrm{WD}=0.439\pm0.002$M$_{\sun}$ and $M_\mathrm{sec}=0.273\pm0.002$M$_{\sun}$ and radii $R_\mathrm{WD}=0.0168\pm0.0003$R$_{\sun}$ and $R_\mathrm{sec}=0.306\pm0.007$R$_{\sun}$. For {\gkvir} we find an inclination of $i=89.5^{\circ}\pm0.6^{\circ}$, masses of $M_\mathrm{WD}=0.564\pm0.014$M$_{\sun}$ and $M_\mathrm{sec}=0.116\pm0.003$M$_{\sun}$ and radii $R_\mathrm{WD}=0.0170\pm0.0004$R$_{\sun}$ and $R_\mathrm{sec}=0.155\pm0.003$R$_{\sun}$. The mass and radius of the white dwarf in {\gkvir} are consistent with evolutionary models for a $50,000$K carbon-oxygen core white dwarf. Although the mass and radius of the white dwarf in {\sdss} are consistent with carbon-oxygen core models, evolutionary models imply that a white dwarf with such a low mass and in a short period binary must have a helium core. The mass and radius measurements are consistent with helium core models but only if the white dwarf has a very thin hydrogen envelope (M$_\mathrm{H}/\mathrm{M}_\mathrm{WD} \leq 10^{-6}$). Such a thin envelope has not been predicted by any evolutionary models. The mass and radius of the secondary star in {\gkvir} are consistent with evolutionary models after correcting for the effects of irradiation by the white dwarf. The secondary star in {\sdss} has a radius $\sim9$ per cent larger than predicted.

\end{abstract}


\section{Introduction}

Detached eclipsing binaries are a primary source of accurate physical properties of stars and stellar remnants. A combination of modelling their light curves and measuring the radial velocities of both components allows us to measure masses and radii to a precision of better than 1 per cent (e.g. \citealt{andersen91}; \citealt{southworth05}; \citealt{southworth07}; \citealt{torres10}). These measurements are crucial for testing theoretical mass-radius relations, which are used in a wide range of astrophysical circumstances such as inferring accurate masses and radii of transiting exoplanets, calibrating stellar evolutionary models and understanding the late evolution of mass transferring binaries such as cataclysmic variables (\citealt{littlefair08}; \citealt{savoury11}). Additionally, the mass-radius relation for white dwarfs has played an important role in estimating the distance to globular clusters \citep{renzini96} and the determination of the age of the galactic disk \citep{wood92}. 

On the one hand, although ubiquitous in the solar neighbourhood, the fundamental properties of low-mass M dwarfs are not as well understood as those of more massive stars \citep{kraus11}. There is disagreement between models and observations, consistently resulting in radii up to 15 per cent larger and effective temperatures 400K or more below theoretical predictions (\citealt{ribas06}; \citealt{lopez07}). These inconsistencies are not only seen in M dwarf eclipsing binaries (\citealt{bayless06}; \citealt{kraus11}) but also in field stars (\citealt{berger06}; \citealt{morales08}) and the host stars of transiting extra-solar planets \citep{torres07}.

\begin{table*}
 \centering
  \caption{Journal of observations. Exposure times for X-shooter observations are for UVB arm, VIS arm and NIR arm respectively. The primary eclipse occurs at phase 1, 2 etc.}
  \label{obs}
  \begin{tabular}{@{}lccccccc@{}}
  \hline
  Date at     &Target    &Instrument&Filter(s)&Start  & Orbital  &Exposure   &Conditions                \\
  start of run&          &          &         &(UT)   & phase    &time (s)   &(Transparency, seeing)    \\
  \hline
  2010/04/05  & {\sdss}  & X-shooter& -       & 00:42 &0.57--0.79&300,338,100&Excellent, $\sim$1 arcsec \\
  2010/04/05  & {\gkvir} & X-shooter& -       & 04:28 &0.12--0.26&300,338,100&Excellent, $\sim$1 arcsec \\
  2010/04/05  & {\sdss}  & X-shooter& -       & 05:57 &0.23--0.45&300,338,100&Excellent, $\sim$1 arcsec \\
  2010/04/05  & {\gkvir} & X-shooter& -       & 07:55 &0.53--0.75&300,338,100&Good, $\sim$1.2 arcsec    \\
  2010/04/21  & {\gkvir} & ULTRACAM &$u'g'i'$ & 07:18 &0.90--1.05&3.0        &Good, $\sim$1.2 arcsec    \\
  2010/04/22  & {\sdss}  & ULTRACAM &$u'g'i'$ & 02:26 &0.40--0.55&2.9        &Good, $\sim$1.2 arcsec    \\
  2010/04/22  & {\gkvir} & ULTRACAM &$u'g'i'$ & 04:14 &0.45--0.55&3.0        &Excellent, $\sim$1 arcsec \\
  2010/04/23  & {\sdss}  & ULTRACAM &$u'g'i'$ & 02:46 &0.40--0.55&2.9        &Excellent, $<$1 arcsec    \\
  2010/04/23  & {\gkvir} & ULTRACAM &$u'g'i'$ & 04:45 &0.40--0.55&3.0        &Excellent, $<$1 arcsec    \\
  2010/04/24  & {\sdss}  & ULTRACAM &$u'g'i'$ & 22:57 &0.90--1.15&2.9        &Good, $\sim$1.2 arcsec    \\
  2010/04/25  & {\gkvir} & ULTRACAM &$u'g'i'$ & 06:29 &0.40--0.55&3.0        &Variable, 1-2 arcsec      \\
  2010/04/29  & {\gkvir} & SOFI     &$J$      & 01:16 &0.40--0.55&15.0       &Excellent, $<$1 arcsec    \\
  2010/04/30  & {\gkvir} & SOFI     &$J$      & 02:17 &0.45--0.60&15.0       &Good, $\sim$1.2 arcsec    \\
  2010/04/30  & {\sdss}  & SOFI     &$J$      & 03:56 &0.40--0.55&10.0       &Excellent, $\sim$1 arcsec \\
  2010/04/30  & {\gkvir} & SOFI     &$J$      & 06:16 &0.90--1.05&15.0       &Excellent, $\sim$1 arcsec \\
  2010/05/01  & {\gkvir} & SOFI     &$J$      & 02:42 &0.40--0.55&15.0       &Good, $\sim$1.2 arcsec    \\
  2011/04/02  & {\sdss}  & SOFI     &$J$      & 01:07 &0.43--0.56&10.0       &Excellent, $<$1 arcsec    \\
  2011/04/02  & {\gkvir} & SOFI     &$J$      & 04:05 &0.38--0.59&15.0       &Excellent, $<$1 arcsec    \\
  2011/04/02  & {\gkvir} & SOFI     &$J$      & 08:02 &0.86--1.05&15.0       &Excellent, $<$1 arcsec    \\
  2011/04/03  & {\sdss}  & SOFI     &$J$      & 01:04 &0.40--0.61&10.0       &Excellent, $\sim$1 arcsec \\
  2011/04/03  & {\gkvir} & SOFI     &$J$      & 05:03 &0.40--0.61&15.0       &Excellent, $\sim$1 arcsec \\
  2011/04/04  & {\sdss}  & SOFI     &$J$      & 01:34 &0.44--0.62&10.0       &Excellent, $\sim$1 arcsec \\
  2011/04/04  & {\gkvir} & SOFI     &$J$      & 05:33 &0.37--0.66&15.0       &Excellent, $\sim$1 arcsec \\
  2011/04/05  & {\sdss}  & SOFI     &$J$      & 01:04 &0.35--0.57&10.0       &Excellent, $\sim$1 arcsec \\
  2011/04/05  & {\gkvir} & SOFI     &$J$      & 05:56 &0.95--1.20&15.0       &Excellent, $\sim$1 arcsec \\
  2011/04/05  & {\sdss}  & SOFI     &$J$      & 05:06 &0.85--1.06&10.0       &Excellent, $\sim$1 arcsec \\
  2011/04/05  & {\gkvir} & SOFI     &$J$      & 06:53 &0.43--0.69&15.0       &Excellent, $\sim$1 arcsec \\
  2011/04/06  & {\sdss}  & SOFI     &$J$      & 01:54 &0.43--0.60&10.0       &Excellent, $<$1 arcsec    \\
  2011/04/06  & {\gkvir} & SOFI     &$J$      & 03:19 &0.90--1.17&15.0       &Excellent, $<$1 arcsec    \\
  2011/04/06  & {\sdss}  & SOFI     &$J$      & 05:59 &0.94--1.12&10.0       &Excellent, $<$1 arcsec    \\
  2011/04/06  & {\gkvir} & SOFI     &$J$      & 07:32 &0.42--0.62&15.0       &Excellent, $<$1 arcsec    \\
\hline
\end{tabular}
\end{table*}

On the other hand, the mass-radius relation for white dwarfs is all but untested observationally. \citet{provencal98} used \emph{Hipparcos} parallaxes to determine the radii for white dwarfs in visual binaries, common proper-motion (CPM) systems and field white dwarfs. However, the radius measurements for all of these systems still rely to some extent on model atmosphere calculations. For field white dwarfs the mass determinations are also indirect. \citet{barstow05} used {\it Hubble Space Telescope}/STIS spectra to measure the mass of Sirius B to high precision, however, their radius constraint still relied on model atmosphere calculations and is therefore less direct when it comes to testing white dwarf mass-radius relations. Double white dwarf eclipsing binaries potentially allow extremely precise measurements of white dwarf masses and radii but have only recently been discovered (\citealt{steinfadt10}; \citealt{parsons11a}; \citealt{brown11}; \citealt{vennes11b}). To date, only a handful of white dwarfs have had their masses and radii model-independently measured, V471 Tau \citep{obrien01}, NN\,Ser \citep{parsons10a}, SDSS J0857+0342 \citep{parsons11b} and SDSS J1210+3347 \citep{pyrzas11}. All of these systems are short period post common envelope binaries (PCEBs), demonstrating the potential of these systems for testing mass-radius relations of both low mass stars and white dwarfs.

In this paper we combine X-shooter spectroscopy, ULTRACAM high-speed photometry and SOFI near-infrared photometry to determine precise system parameters for the eclipsing PCEBs {\SDSS} (henceforth {\sdss}) \citep{nebot09} and {\gkvir} \citep{green78}. We then compare our mass and radius measurements with theoretical mass-radius relations for white dwarfs and low-mass stars.

\section{Observations and their reduction}

\subsection{ULTRACAM photometry}

{\gkvir} and {\sdss} were observed with ULTRACAM mounted as a visitor instrument on the 3.5m New Technology Telescope (NTT) at La Silla in April 2010. ULTRACAM is a high-speed, triple-beam CCD camera \citep{dhillon07} which can acquire simultaneous images in three different bands; for our observations we used the SDSS $u'$, $g'$ and $i'$ filters.  A complete log of these observations is given in Table~\ref{obs}. The data collected for {\gkvir} were combined with previous observations of the system taken with ULTRACAM (see \citealt{parsons10b} for details of these observations). We windowed the CCD in order to achieve exposure times of $\sim$3 seconds which we varied to account for the conditions. The dead time between exposures was $\sim 25$ ms.

\begin{table*}
 \centering
  \caption{Comparison star apparent magnitudes and coordinates. Magnitudes listed are for those bands for which the comparison star was used.}
  \label{comps}   
  \begin{tabular}{@{}lcccccc@{}}
  \hline
  Target   & Comp      & Comp      & $u'$ & $g'$ & $i'$ & $J$  \\
  star     & RA        & Dec       &      &      &      &      \\
  \hline
  {\gkvir} &14:15:22.86&+01:19:12.7& 14.5 & 13.5 & 13.1 &   -  \\
  {\sdss}  &12:12:54.97&-01:20:25.0& 13.8 & 12.5 & 11.9 &   -  \\
  {\gkvir} &14:15:29.31&+01:17:37.9&   -  &   -  &   -  & 13.7 \\
  {\sdss}  &12:13:00.51&-01:23:23.3&   -  &   -  &   -  & 13.4 \\
\hline
\end{tabular}
\end{table*}

All of these data were reduced using the ULTRACAM pipeline software. Debiassing, flatfielding and sky background subtraction were performed in the standard way. The source flux was determined with aperture photometry using a variable aperture, whereby the radius of the aperture is scaled according to the full width at half maximum (FWHM). Variations in observing conditions were accounted for by determining the flux relative to a comparison star in the field of view. Apparent magnitudes and coordinates for each of the comparison stars used are given in Table~\ref{comps}. We flux calibrated our targets by determining atmospheric extinction coefficients in each of the bands in which we observed and calculated the absolute flux of our targets using observations of standard stars (from \citealt{smith02}) taken in twilight. Using our absorption coefficients we extrapolated all fluxes to an airmass of $0$. The systematic error introduced by our flux calibration is  $<0.1$ mag in all bands.

\subsection{SOFI $J$-band photometry}

We observed both {\gkvir} and {\sdss} with SOFI \citep{moorwood98} mounted at the NTT in April 2010 and April 2011. The observations were made in fast photometry mode equipped with a $J$-band filter. We windowed the detector to achieve a cycle time of $\sim10$--$15$ seconds and offset the telescope every 10 minutes in order to improve sky subtraction. A summary of these observations is given in Table~\ref{obs}.

The dark current removal (which also removes the bias) and flatfielding were performed in the standard way. Sky subtraction was achieved by using observations of the sky when the target had been offset. The average sky level was then added back so that we could determine the source flux and its uncertainty with standard aperture photometry, using a variable aperture, within the ULTRACAM pipeline. A comparison star was used to account for variations in observing conditions, details of these are given in Table~\ref{comps}. Flux calibration was done using the comparison star $J$-band magnitude retrieved from the 2MASS catalogue \citep{skrutskie06}. 

\subsection{X-shooter spectroscopy}

{\gkvir} and {\sdss} were both observed with X-shooter \citep{dodorico06} mounted at the VLT-UT2 telescope on the night of the 4th of April 2010. Due to the long orbital periods of both systems ($\sim 8$ hours) we targeted the quadrature phases, since these phases are the most sensitive to the radial velocity amplitude. Conditions throughout the night were excellent with seeing consistently below 1 arcsec. Details of these observations are listed in Table~\ref{obs}.

X-shooter is a medium resolution spectrograph consisting of 3 independent arms that give simultaneous spectra longward of the atmospheric cutoff (0.3 microns) in the UV (the ``UVB'' arm), optical (the ``VIS'' arm) and up to 2.5 microns in the near-infrared (the ``NIR''arm). We used slit widths of 0.8'', 0.9'' and 0.9'' in X-shooter's three arms and binned by a factor of two in the dispersion direction, resulting in a resolution of $R \sim 7,000$. We used exposure times of 300 seconds in the UVB arm, 338 seconds in the VIS arm and $3 \times 100$ seconds in the NIR arm. After each exposure we nodded along the slit to help the sky subtraction in the NIR arm.

\begin{figure*}
\begin{center}
 \includegraphics[width=0.97\textwidth]{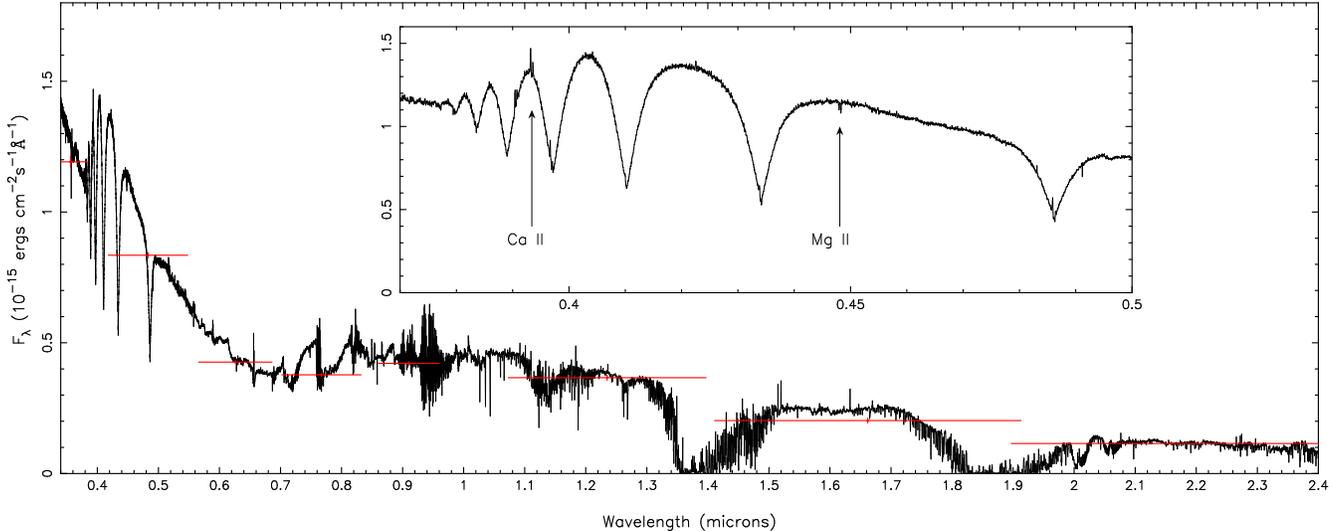}
 \vspace{2.5mm}
 \caption{Averaged X-shooter spectrum of {\sdss}. The SDSS $u'g'r'i'z'$ and 2MASS $JHK$ magnitudes and filter widths are also shown. A zoom in on the white dwarf features are shown inset with the narrow {\caii} and {\mgii} absorption features labelled, although the {\caii} feature is somewhat filled in by emission from the secondary star.}
 \label{sdss1212_spec}
 \end{center}
\end{figure*}

The reduction of the raw frames was conducted using the standard pipeline release of the X-shooter Common Pipeline Library (CPL) recipes (version 1.3.7) within ESORex, the ESO Recipe Execution Tool, version 3.9.0. The standard recipes were used to optimally extract and wavelength calibrate each spectrum. For the NIR arm we combined frames taken at different nod positions to improve the sky subtraction, however, this does result in a reduction in orbital phase resolution. The instrumental response was removed by observing the spectrophotometric standard star GJ 440 and dividing it by a flux table of the same star \citep{hamuy92} to produce the response function, this was also used to apply a telluric correction to the spectra. We then heliocentrically corrected the wavelength scales of each of the spectra. We achieved a signal-to-noise ($S/N$) for {\gkvir} of $\sim20$ in the UVB arm per exposure, $\sim15$ in the VIS arm per exposure and $\sim5$ in the NIR arm per pair of nodded exposures. For {\sdss} we achieved a $S/N$ of $\sim20$ in the UVB arm per exposure, $\sim30$ in the VIS arm per exposure and $\sim20$ in the NIR arm per pair of nodded exposures.

\subsection{Flux calibration}

The ULTRACAM and SOFI light curves were used to flux calibrate the X-shooter spectra. We fitted a model to each of the light curves (see Section~\ref{lcurvemodel}) in order to reproduce the light curve as closely as possible. The model was then used to predict the flux at the times of each of the X-shooter observations ({\gkvir} shows no stochastic variations or flaring, whilst small flares were seen in one ULTRACAM observation and two SOFI observations of {\sdss} but were removed before fitting).

We then derived synthetic fluxes from the spectra for the ULTRACAM $u'$, $g'$, $r'$, $i'$ and $z'$ filters as well as the SOFI $J$, $H$ and $K$ filters. We extrapolated the light curve models to those bands not covered by our photometry. We then calculated the difference between the model and synthetic fluxes and fitted a second-order polynomial to them. This correction was then applied to each spectrum. This corrects for variable extinction across the wavelength range, as well as variations in seeing.

\section{results}

\subsection{{\sdss}}
{\sdss} was initially listed as a quasar candidate from the Sloan Digital Sky Survey (SDSS) by \citet{richards04}. \citet{silvestri06} reclassified it as a white dwarf plus main-sequence binary and eclipses were discovered by \citet{nebot09} who derived the basic system parameters. However, their analysis was limited by the fact that they did not resolve the white dwarf ingress or egress and could not measure the radial velocity amplitude of the white dwarf. They found that {\sdss} contained a relatively hot ($17,700\pm300$K) low mass white dwarf with an active M4 main-sequence companion in an $8^\mathrm{h}3^\mathrm{m}$ period.

\subsubsection{Spectral features}
Figure~\ref{sdss1212_spec} shows an average spectrum of {\sdss}. The white dwarf dominates the spectrum at wavelengths shorter than 0.55 microns whilst at longer wavelengths the spectral features of the secondary star dominate. There are also numerous emission lines throughout the spectrum originating from the secondary star. Additionally, both {\caii} 3934{\AA} and {\mgii} 4481{\AA} absorption from the white dwarf are seen, likely the result of low level accretion from the wind of the secondary star. The fact that absorption features are seen from both stars allows us to measure the radial velocities for both components of the system directly.

\subsubsection{Atmospheric parameters of the white dwarf} \label{wdtemp}
\citet{nebot09} determined the temperature of the white dwarf in {\sdss} by decomposing the SDSS spectrum. We computed the average X-shooter spectrum of the white dwarf by removing the M-dwarf contribution from the individual spectra using observations of the M4 star GJ 447, and shifting the residual spectra to the rest frame of the white dwarf, however this does not remove the emission components. We then fit the average white dwarf spectrum following the method outlined in \citet{rebassa07,rebassa10}, using a grid of pure hydrogen model atmospheres calculated with the code described by \citet{koester10}. We down-weight the cores of the Balmer lines since they are contaminated by emission from the secondary star, we do not use the H$\alpha$ line since the secondary star dominates at this wavelength, all other lines were used. We find a temperature of $17,707\pm35$K, a surface gravity of $\log{g}=7.51\pm0.01$ and a distance of $228\pm5$pc, all of which is consistent with the results from \citet{nebot09} (and from our light curve fit, see Section~\ref{sec:discuss}). Note that these are purely statistical uncertainties.

\begin{figure}
  \begin{center}
    \includegraphics[width=\columnwidth]{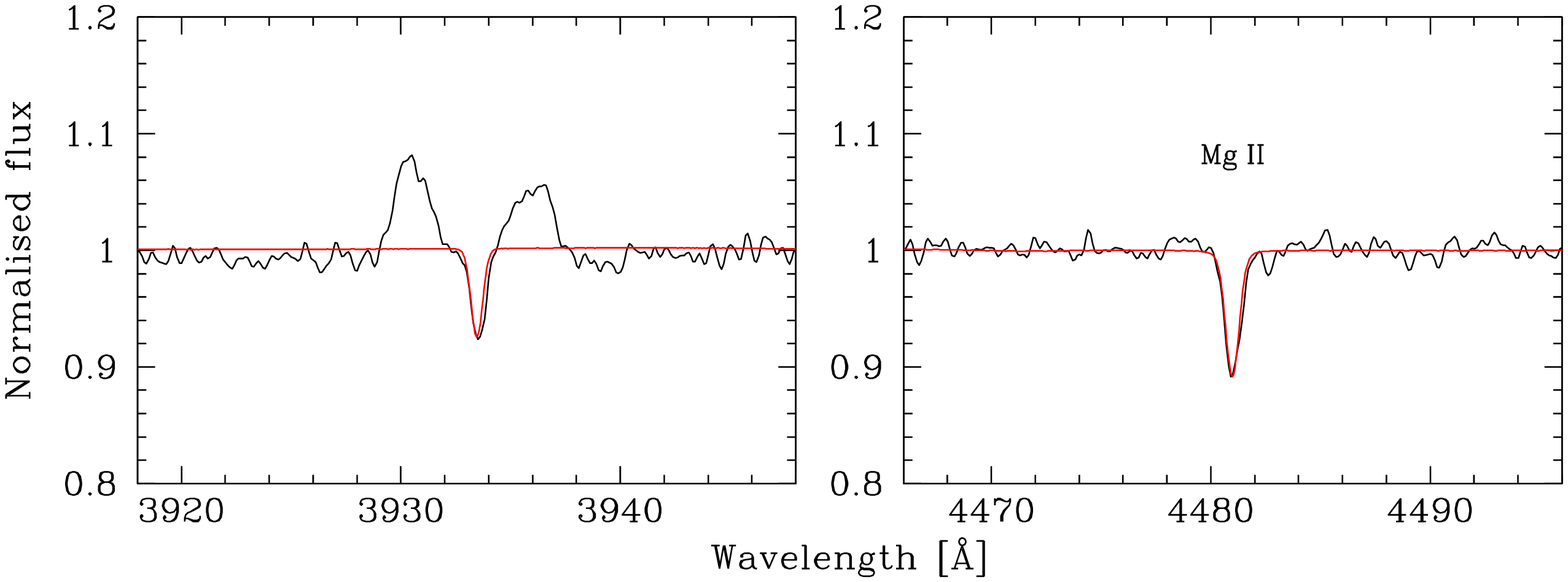}
    \caption{The X-shooter spectra of {\sdss}, averaged in the white dwarf rest-frame and normalised (black) along with the best-fit white dwarf model ($\Teff=17707$, $\log{g}=7.51$) and $\log[\mathrm{Mg/H}]=-5.8$. The Ca\,K line is contaminated by the emission from the secondary star (which appears shifted/smeared in the rest frame of the white dwarf), however, adopting the same abundance for Ca provides a reasonable match to the photospheric absorption line.}
  \label{f-mgfit}
  \end{center}
\end{figure}

Next, we fitted the equivalent width of the observed \mgii\,4481{\AA} absorption line ($90\pm20$\,m{\AA}) by varying the Mg abundance in the synthetic spectrum, keeping $\Teff$ and $\log{g}$ fixed to the values determined above. The best-fit abundance by number is $\log[\mathrm{Mg/H}]=-5.8\pm0.1$, corresponding to $\simeq4$\% of the solar value. Figure~\ref{f-mgfit} shows the fit to the \mgii\,4481{\AA} line. The Ca\,K line is significantly contaminated by the emission line of the companion star, however, Figure~\ref{f-mgfit} illustrates that adopting the Mg abundance also for Ca matches the observed Ca\,K line reasonably well.

Since there is no convection zone acting as a reservoir for the accreted elements, the settling times vary throughout the atmosphere. At optical depth $\sim2/3$, representative for the visible spectrum, the timescale for Mg is a few tens of days.  The diffusion time scale for the temperature and surface gravity derived above is $\sim2$\,months \citep{koester06}, it is hence plausible to assume accretion-diffusion equilibrium. Hence, $X_\mathrm{Mg} \rho v_\mathrm{Mg} = \mathrm{constant}$, with $X_\mathrm{Mg}$ and $v_\mathrm{Mg}$ the mass fraction and the diffusion velocity (relative to hydrogen) of Mg, and $\rho$ the mass density. We compute at $\tau_\mathrm{Ross} \approx 2/3$ $\rho=5.23\times10^{-8}\mathrm{g\,cm^{-3}}$, $X_\mathrm{Mg}=3.76\times10^{-5}$, and $v_\mathrm{Mg}=8.48\times10^{-2}\mathrm{cm\,s^{-1}}$, i.e. a mass flux of $1.67\times10^{-13}\mathrm{g\,cm^{-2}s^{-1}}$. Multiplying by the white dwarf surface area, with $R_\mathrm{WD}$ from Table~\ref{sys_para}, the total mass flux of Mg is $2.9\times10^{6}\mathrm{g\,s^{-1}}$~--~which is equal to the mass accretion rate of Mg at the surface of the white dwarf. Assuming that the donor star transfers material of solar composition, the total accretion rate is then obtained by dividing the Mg rate by the mass fraction of Mg in the Sun, i.e. $\dot M\simeq4.0\times10^9\mathrm{g\,s^{-1}}$ or $6.4\times10^{-17}\msy$. This value is right in the middle of the accretion rates calculated for the pre-CVs RR\,Cae, UZ\,Sex, EG\,UMa, LTT\,560 and SDSS\,J121010.1+334722.9 by \citet{debes06}, \citet{tappert11}, and \citet{pyrzas11}, $9\times10^{-19}\msy$ to $5\times10^{-15}\msy$.

\begin{table}
 \centering
  \caption{White dwarf absorption features in {\sdss}.}
  \label{wd_feats_1212}
  \begin{tabular}{@{}lcc@{}}
  \hline
  Line      & $K_\mathrm{WD}$ & $\gamma_\mathrm{WD}$ \\
            & ($\kms$)       & ($\kms$)             \\
 \hline
{\caii}   3933.663 & $102.71\pm1.32$ & $37.36\pm1.12$ \\
H$\delta$ 4101.735 & $106.22\pm2.21$ & $35.14\pm1.85$ \\
H$\gamma$ 4340.465 & $104.28\pm1.24$ & $37.43\pm1.08$ \\
{\mgii}   4481.126 & $104.53\pm0.78$ & $35.75\pm0.67$ \\
H$\beta$  4861.327 & $105.49\pm0.93$ & $35.25\pm0.42$ \\
H$\alpha$ 6562.760 & $103.32\pm1.13$ & $38.66\pm0.97$ \\
\hline
\end{tabular}
\end{table}

\subsubsection{White dwarf radial velocity}
The orbital phases of the X-shooter spectra of {\sdss} were determined using the ephemeris derived in Section~\ref{ephem_1212}. As previously mentioned both {\caii} and {\mgii} absorption from the white dwarf are present. Additionally, the cores of the longer wavelength hydrogen absorption lines (H$\delta$ to H$\alpha$) are narrow and suitable for radial velocity measurements.

\begin{figure*}
  \begin{center}
    \includegraphics[width=0.9\textwidth]{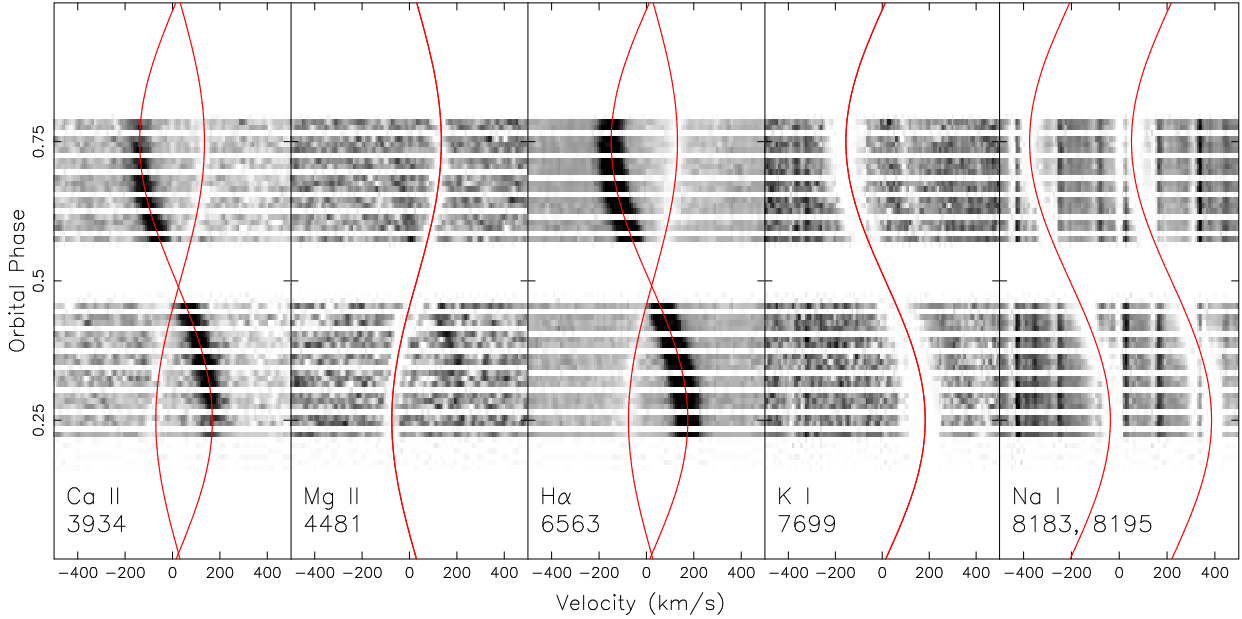}
    \vspace{4mm}
  \caption{Trailed spectra of several lines in {\sdss}. The grey-scale runs from white (75 per cent of the continuum level) to black (125 per cent of the continuum level). The {\caii} 3934{\AA} line shows an absorption component from the white dwarf and an emission component from the M star. The {\mgii} 4481{\AA} line is from the white dwarf, a weak Fe\,{\sc i} emission line is also visible. H$\alpha$ shows both absorption from the white dwarf and emission from the M dwarf. The K\,{\sc i} absorption line and the Na\,{\sc i} absorption doublet originate from the M star, a telluric correction was applied but artifacts still remain in the Na\,{\sc i} trail. The red lines (online version only) show the best fits to the lines.}
  \label{1212_trails}
  \end{center}
\end{figure*}

We measured the radial velocities of the absorption lines by simultaneously fitting all of the spectra. We used a combination of a straight line and Gaussians for each spectrum (including a broad Gaussian component to account for the wings of the absorption in the case of the Balmer lines) and allowed the position of the Gaussians to change velocity according to
\begin{eqnarray}
V = \gamma + K\sin(2 \pi \phi), \nonumber
\end{eqnarray}
for each star, where $\gamma$ is the velocity offset of the line from its rest wavelength and $\phi$ is the orbital phase of the spectrum. 

The parameters determined from the fits to the white dwarf absorption features are listed in Table~\ref{wd_feats_1212}. The fits to several of the absorption features are also shown in Figure~\ref{1212_trails}, the fit to the H$\alpha$ line is also shown in Figure~\ref{sdss1212_rvs}. Taking a weighted average of the radial velocities gives a radial velocity amplitude of the white dwarf of $K_\mathrm{WD} = 104.4\pm0.5\,\kms$.

\subsubsection{Secondary star radial velocity}
\label{sdss1212_k2}
There are both absorption and emission features originating from the secondary star seen in the X-shooter spectra. However, the emission features are due to irradiation from the white dwarf hence they do not track the true radial velocity of the secondary star but the centre of light of the emission region, which will be offset towards the white dwarf. Therefore we can only directly measure the velocity amplitude for the secondary star via the absorption lines.

\begin{figure}
  \begin{center}
    \includegraphics[width=\columnwidth]{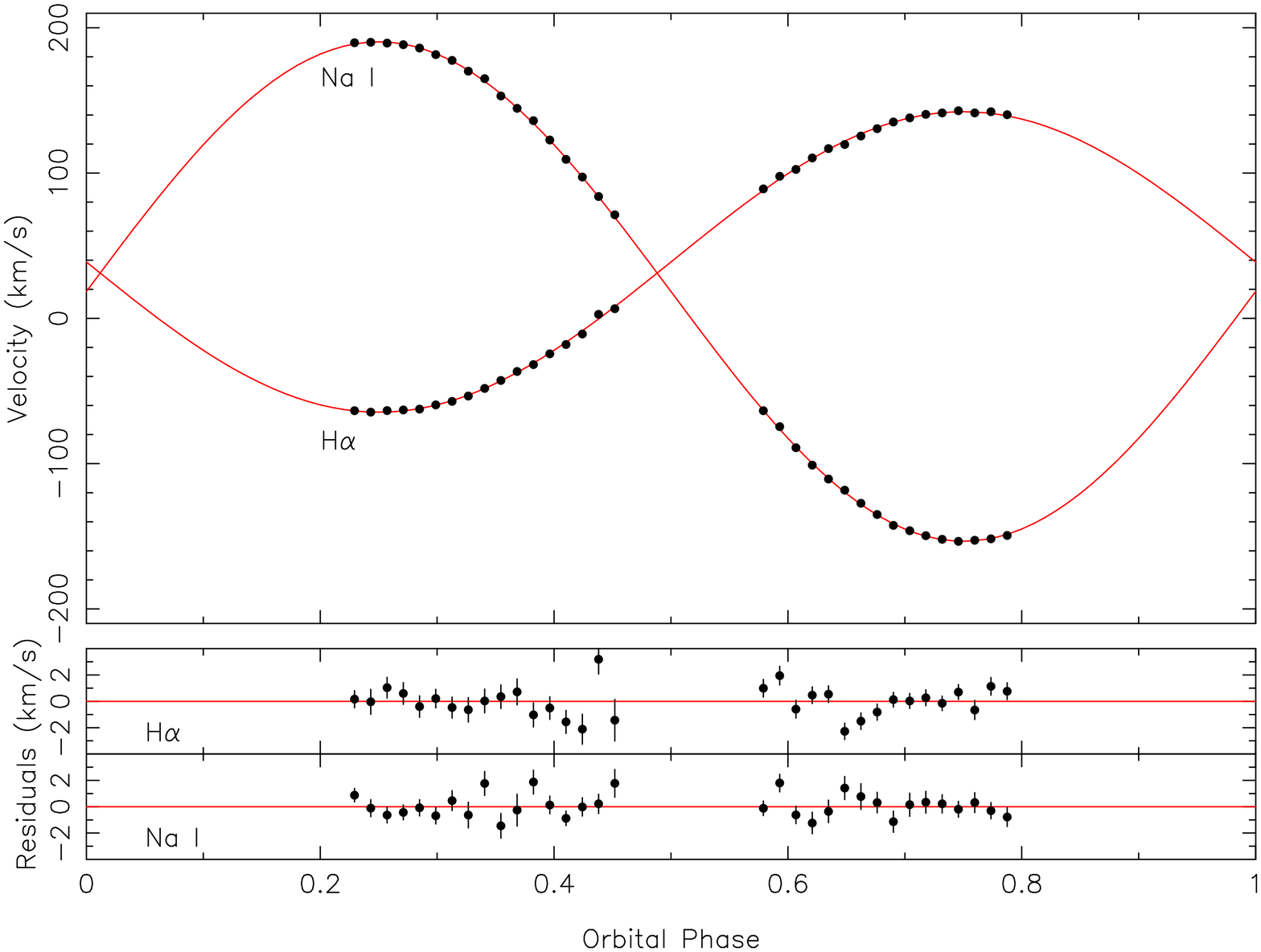}
    \vspace{1mm}
  \caption{Radial velocity fits to the H$\alpha$ absorption from the white dwarf and Na\,{\sc i} 8183{\AA} absorption from the secondary star in {\sdss} with residuals plotted below.}
  \label{sdss1212_rvs}
  \end{center}
\end{figure}

\begin{table}
 \centering
  \caption{Secondary star atomic absorption features in {\sdss}. There are several additional atomic absorption features in the NIR arm spectra but the $S/N$ of these lines are too low to reliably fit them.}
  \label{sec_feats_1212}
  \begin{tabular}{@{}lcc@{}}
  \hline
  Line      & $K_\mathrm{sec}$ & $\gamma_\mathrm{sec}$ \\
            & ($\kms$)       & ($\kms$)             \\
 \hline
K\,{\sc i} 7664.899  & $169.36\pm1.13$ & $18.27\pm0.97$ \\
K\,{\sc i} 7698.964  & $169.04\pm0.83$ & $19.28\pm0.71$ \\
Na\,{\sc i} 8183.256 & $168.77\pm0.45$ & $18.44\pm0.38$ \\
Na\,{\sc i} 8194.824 & $167.42\pm0.48$ & $19.52\pm0.41$ \\
Na\,{\sc i} 11381.45 & $165.24\pm3.86$ & $21.51\pm3.21$ \\
Na\,{\sc i} 11403.78 & $166.87\pm1.20$ & $19.14\pm1.07$ \\
K\,{\sc i} 11690.219 & $170.22\pm1.37$ & $20.86\pm1.24$ \\
K\,{\sc i} 12432.274 & $168.63\pm1.01$ & $19.47\pm0.88$ \\
K\,{\sc i} 12522.141 & $167.40\pm1.03$ & $21.48\pm0.72$ \\
\hline
\end{tabular}
\end{table}

Sodium and Potassium absorption lines are present in the spectra which we fitted in the same way as the white dwarf absorption features. The result of these fits are given in Table~\ref{sec_feats_1212}. There were also a number of molecular absorption features which we do not fit due to the uncertainty in their rest wavelengths and their broad, non-Gaussian profiles. Other atomic absorption features are seen at longer wavelengths (for example: the Na\,{\sc i} doublet at 2.2 microns) but the $S/N$ of these features are too low to reliably fit them. A weighted average of the radial velocities of the lines in Table~\ref{sec_feats_1212} gives a radial velocity amplitude of the secondary star of $K_\mathrm{sec}=168.3\pm0.3\,\kms$. This is somewhat lower than the value of $K_\mathrm{sec}=181\pm3\,\kms$ found by \citet{nebot09}. To try and resolve this difference we re-fitted the radial velocity data from \citet{nebot09} and found a $\chi^2=25$ for 10 points with 2 variables, implying that they slightly underestimated their error. Accounting for this, their value is consistent with ours to within $2.5\sigma$ but, based on our higher resolution and additional clean features (not affected by telluric absorption), we favour our value for $K_\mathrm{sec}$. We find no evidence that the absorption lines are affected by irradiation from the white dwarf. The equivalent widths of these lines do not vary with phase and no effects are visible in the radial velocity curves.

We also measured the radial velocity amplitudes of all of the identified emission lines. However, for these lines we allow the height of the emission line to vary with orbital phase according to
\begin{eqnarray}
H = H_0 - H_1\cos(2 \pi \phi), \nonumber
\end{eqnarray}
which allows the height to peak at phase 0.5, where the irradiation effect is largest. This approach gives better fits than keeping the height at a fixed value. 

The results of these fits are given in Table~\ref{sec_emis_1212}, in the appendix. We find that the emission lines give a range of radial velocities, likely due to the different optical depths of the lines \citep{parsons10a}. Figure~\ref{1212_trails} shows the fits to the Na\,{\sc i} doublet as well as several emission lines. The fit to the Na\,{\sc i} 8183{\AA} line is shown in Figure~\ref{sdss1212_rvs}.

\subsubsection{White dwarf's gravitational redshift}
\label{rshift_section}

General relativity tells us that the gravitational redshift of a white dwarf is given by
\begin{eqnarray}
\label{redshift}
V_z = 0.635 (M/M_{\sun})(R_{\sun}/R)\, \kms
\end{eqnarray}
where $M$ and $R$ are the mass and radius of the white dwarf. Furthermore, if we know the radial velocity amplitudes of the two stars then Kepler's third law tells us
\begin{eqnarray}
\label{keplers}
M_\mathrm{WD} = \frac{P K_\mathrm{sec} (K_\mathrm{WD}+K_\mathrm{sec})^2}{2 \pi G \sin^3{i}}
\end{eqnarray}
where $P$ is the orbital period, and $i$ is the orbital inclination. Therefore for a given inclination we can calculate the mass of the white dwarf via Eq (\ref{keplers}) and the radius of the white dwarf via a model fitted to the primary eclipse, and thus predict a redshift. Hence we can use the measurement of the gravitational redshift to constrain the inclination by rejecting light curve models which do not satisfy this constraint.

The gravitational redshift can be measured from the difference in the velocities of the two components ($\gamma_\mathrm{WD}-\gamma_\mathrm{sec}$). $\gamma_\mathrm{sec}$ is tightly constrained due to the large number of emission lines as well as many absorption features. Taking an inverse variance weighted mean of the secondary star line velocities (from Tables~\ref{sec_feats_1212} and \ref{sec_emis_1212}) we found $\gamma_\mathrm{sec} = 19.93\pm0.06\,\kms$. We calculated $\gamma_\mathrm{WD}$ in the same way using the values in Table~\ref{wd_feats_1212} which gave $\gamma_\mathrm{WD} = 36.0\pm0.3\,\kms$. We found no evidence of pressure shifts in these lines, although these are expected to be small for calcium and magnesium \citep{vennes11a}.

Using these measurements we determined the gravitational redshift of the white dwarf to be $V_z = 16.1\pm0.3\,\kms$. The true gravitational redshift of the white dwarf will actually be slightly higher than this value since the measured value includes the effects of the secondary star (see Section~\ref{lcurvemodel} for details of these corrections).

\subsubsection{Eclipse time}
\label{ephem_1212}

\begin{table}
 \centering
  \caption{Eclipse times for {\sdss}. (1) \citet{nebot09}, (2) This paper.}
  \label{sdss1212_times}
  \begin{tabular}{@{}lcc@{}}
  \hline
Cycle & MJD(BTDB)     & Reference \\ 
No.   & (mid-eclipse) &           \\
 \hline
0    & 54104.2092(21)    & (1) \\
122  & 54145.1854(8)     & (1) \\
125  & 54146.1929(8)     & (1) \\
205  & 54173.0628(10)    & (1) \\
410  & 54241.9170(21)    & (1) \\
1455 & 54592.9008(14)    & (1) \\
3593 & 55310.9934268(59) & (2) \\
\hline
\end{tabular}
\end{table}

We recorded one eclipse of {\sdss} with ULTRACAM. This provides the first high-precision eclipse time for this system. The SOFI eclipses are not suitable for long term period studies as they are not precisely timed. The eclipse time, as well as all previous eclipse times, are listed in Table~\ref{sdss1212_times}. The new time is consistent with the ephemeris of \citet{nebot09} but since it is of higher precision we use it to update the ephemeris to 
\begin{eqnarray}
\mathrm{MJD(BTDB)} & = & 54104.209\,17(48) +\, 0.335\,870\,93(13) E, \nonumber
\end{eqnarray}
where we have used barycentric dynamical time (TDB) corrected for the light travel time to the barycentre of the solar system (BTDB). Since this is the first precise eclipse time, no long-term period trend is yet visible in the data.
 
\subsubsection{Spectral type of the secondary star}

From their spectral decomposition \citet{nebot09} determined the spectral type of the secondary star in {\sdss} to be M$4\pm1$. We detect the secondary star during the eclipse in all bands except the $u'$ band giving us multi-colour information. We measure magnitudes for the secondary star of $g'=19.73\pm0.04$, $i'=17.384\pm0.004$ and $J=14.949\pm0.001$. The $i'-J=2.435\pm0.004$ colour implies a spectral type of M4 for the secondary star (\citealt{hawley02}, Table~3), consistent with the result from \citet{nebot09}. The distance determined from the spectroscopic fit ($228\pm5$ pc) means that the absolute $J$ band magnitude of the secondary star is $M_J=8.2$. Using the $M_J$-spectral type relation from \citet{hawley02}, also gives a spectral type of M4. 

\subsubsection{Modelling the light curves}
\label{lcurvemodel}

\begin{figure*}
  \begin{center}
    \includegraphics[width=0.9\textwidth]{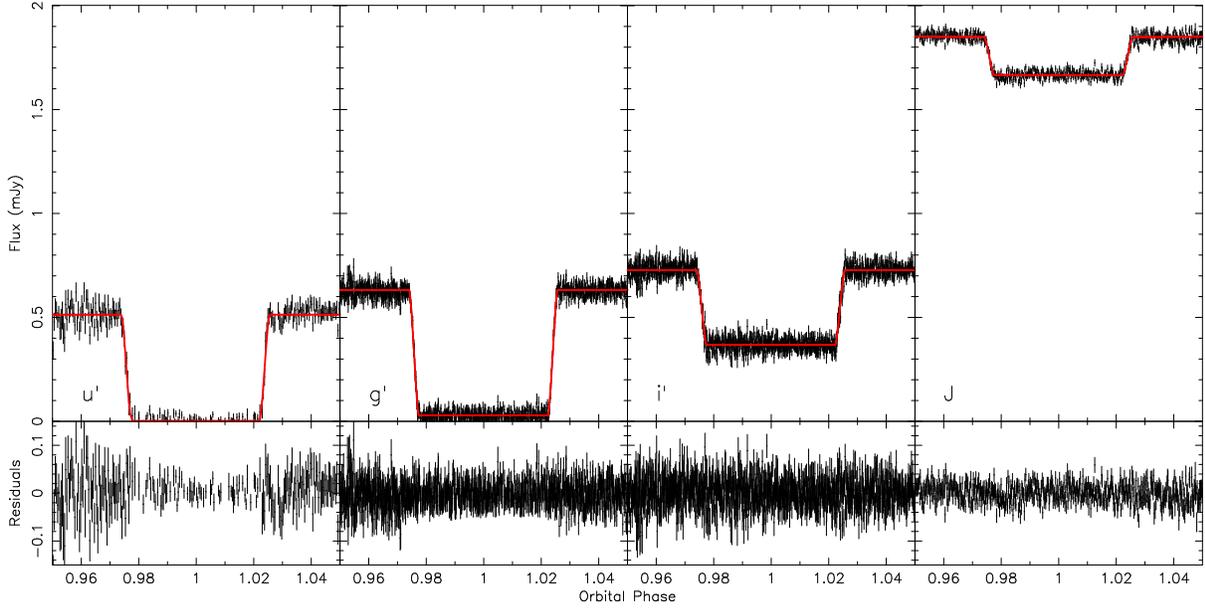}
    \vspace{2mm}
  \caption{ULTRACAM $u'$, $g'$ and $i'$ and SOFI $J$ band primary eclipses of {\sdss} with model fits and residuals.}
  \label{sdss1212_lcurves}
  \end{center}
\end{figure*}

\begin{table*}
 \centering
  \caption{Parameters from Markov chain Monte Carlo minimisation for {\sdss}, some fitted, some fixed a priori (those without quoted uncertainties). $a$ and $b$ are the quadratic limb darkening coefficients. $A$ is the fraction of the irradiating flux from the white dwarf absorbed by the secondary star.}
  \label{sdss1212_lcfit}
  \begin{tabular}{@{}lcccc@{}}
  \hline
 Parameter & $u'$ & $g'$ & $i'$ & $J$ \\
 \hline 
$i$ (deg)          & $86.1\pm2.2$      & $85.7\pm0.6$      & $85.8\pm0.8$      & $85.5\pm0.8$      \\
$r_\mathrm{WD}/a$    & $0.0092\pm0.0002$ & $0.0092\pm0.0002$ & $0.0093\pm0.0002$ & $0.0092\pm0.0002$ \\
$r_\mathrm{sec}/a$   & $0.171\pm0.016$   & $0.171\pm0.005$   & $0.170\pm0.006$   & $0.172\pm0.006$   \\
$T_\mathrm{eff,sec}$ (K) & $2618\pm112$      & $2947\pm26$       & $3009\pm41$       & $3342\pm42$       \\
$a_\mathrm{WD}$      & $0.2444$          & $0.1340$          & $0.1071$          & $0.0639$          \\
$b_\mathrm{WD}$      & $0.2256$          & $0.2899$          & $0.1891$          & $0.1342$          \\
$a_\mathrm{sec}$     & $0.5866$          & $0.6720$          & $0.4193$          & $0.0254$          \\
$b_\mathrm{sec}$     & $0.2959$          & $0.2660$          & $0.4109$          & $0.4826$          \\
$A$                & $1.60\pm0.50$     & $0.81\pm0.09$     & $0.45\pm0.03$     & $0.39\pm0.02$     \\
 \hline
\end{tabular}
\end{table*} 

The light curves of {\sdss} are characterised by a deep eclipse of the white dwarf (which gets shallower in the longer wavelength bands). There is little variation out of eclipse and we found no evidence of a secondary eclipse in any of the light curves.

We fitted all of our photometry (for both {\sdss} and {\gkvir}) using a code written to produce models for the general case of binaries containing a white dwarf (see \citealt{copperwheat10} for details). It has been used in the study of other white dwarf-main sequence binaries (\citealt{pyrzas09}; \citealt{parsons10a}). The program subdivides each star into small elements with a geometry fixed by its radius as measured along the direction of centres towards the other star. Roche geometry distortion and irradiation of the secondary star are included, the irradiation is approximated by $\sigma T^{\prime}{}_\mathrm{sec}^{4}= \sigma T_\mathrm{sec}^{4}+ A F_\mathrm{irr}$ where $T^{\prime}{}_\mathrm{sec}$ is the modified temperature and $T_\mathrm{sec}$ is the temperature of the unirradiated main-sequence star, $\sigma$ is the Stefan-Boltzmann constant, $A$ is the fraction of the irradiating flux from the white dwarf absorbed by the secondary star and $F_\mathrm{irr}$ is the irradiating flux, accounting for the angle of incidence and distance from the white dwarf. 

The parameters needed to define the model were: the mass ratio, $q = M_\mathrm{sec}/M_\mathrm{WD}$, the inclination, $i$, the sum of the unprojected stellar orbital speeds $V_s=(K_\mathrm{WD}+K_\mathrm{sec})/\sin(i)$, the stellar radii scaled by the orbital separation $R_\mathrm{sec}/a$ and $R_\mathrm{WD}/a$, the unirradiated temperatures, $T_\mathrm{eff,WD}$ and $T_\mathrm{eff,sec}$, quadratic limb darkening coefficients for the both stars, the time of mid eclipse, $T_{0}$ and the period, $P$. Note that the temperatures are really just flux scaling parameters and only approximately correspond to the actual temperatures.

The light curves are only weakly dependent upon $q$ and $V_s$. However, we can use them, as well as our measurements of the radial velocity amplitudes and the gravitational redshift of the white dwarf, to help constrain the orbital inclination. This is done by computing the radial velocity amplitudes via
\begin{eqnarray}
K_\mathrm{WD}  & = & (q/(1+q)) V_s \sin{i} \\
K_\mathrm{sec} & = & (1/(1+q)) V_s \sin{i},
\end{eqnarray}
we can also compute the masses using
\begin{eqnarray}
M = P V_s^3/2 \pi G, \label{eqn:redshift1}
\end{eqnarray}
where $M$ is the total system mass. The individual masses are then
\begin{eqnarray}
M_\mathrm{WD}  & = & M/(1+q) \\
M_\mathrm{sec} & = & q M_\mathrm{WD}.
\end{eqnarray}
The orbital separation $a$ is then calculated from
\begin{eqnarray}
a = V_s/2 \pi P, \label{eqn:redshift2}
\end{eqnarray}
allowing us to calculate the radii of the two stars. Combining all these calculations yields the masses and radii of both stars. We can then use these to calculate the gravitational redshifts using
\begin{eqnarray}
V_{z,\mathrm{WD}}  & = & 0.635 \left( \frac{M_\mathrm{WD}}{R_\mathrm{WD}} + \frac{M_\mathrm{sec}}{a} \right) + \frac{(K_\mathrm{WD}/\sin{i})^2}{2c} \\
V_{z,\mathrm{sec}} & = & 0.635 \left( \frac{M_\mathrm{sec}}{R_\mathrm{sec}} +\frac{M_\mathrm{WD}}{a} \right) + \frac{(K_\mathrm{sec}/\sin{i})^2}{2c},
\end{eqnarray}
then the value $V_z = V_{z,\mathrm{WD}} - V_{z,\mathrm{sec}}$ is equivalent to the measured redshift from the X-shooter spectra, which takes into account the effects of the secondary star on the measurement of the white dwarf's gravitational redshift. Therefore after we have generated a model we can compute $K_\mathrm{WD}$, $K_\mathrm{sec}$ and $V_z$ from the fitted parameters and either reject or accept that model based on how close they are to the measured values.

\begin{figure*}
\begin{center}
 \includegraphics[width=0.97\textwidth]{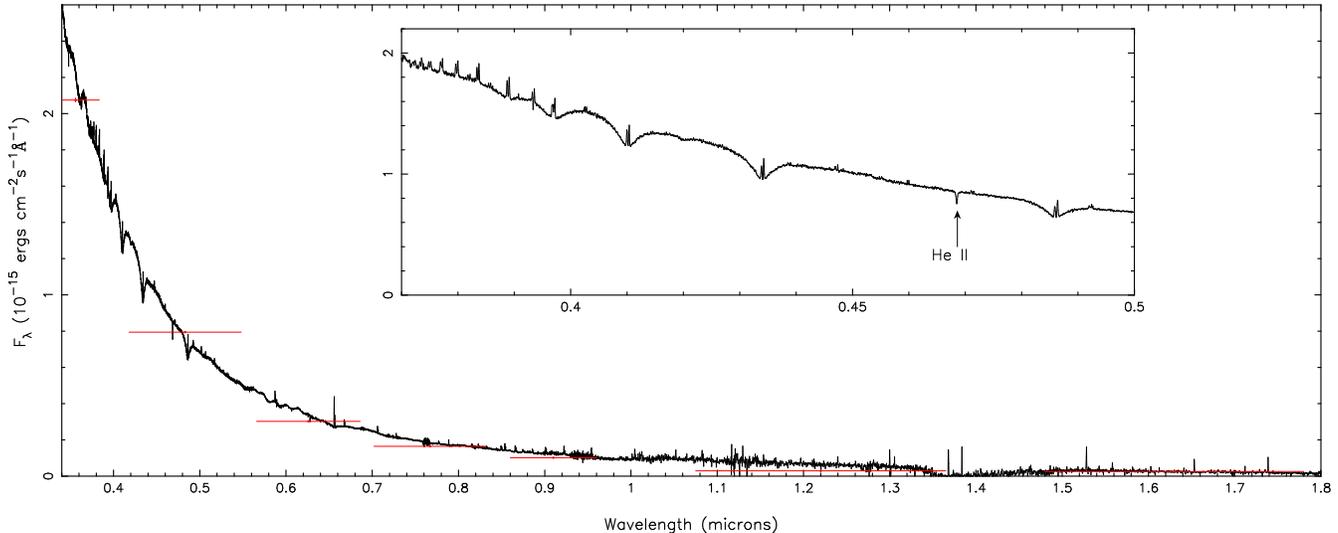}
 \vspace{2.5mm}
 \caption{Averaged X-shooter spectrum of {\gkvir}. The SDSS $u'g'r'i'z'$ and UKIDSS $JH$ magnitudes and filter widths are also shown, the UKIDSS $J$ band observations were made during the primary eclipse. A zoom in on the white dwarf features are shown inset with the narrow {\heii} absorption feature labelled.}
 \label{gkvir_spec}
 \end{center}
\end{figure*}

We used the Markov Chain Monte Carlo (MCMC) method to determine the distributions of our model parameters \citep{press07}. The MCMC method involves making random jumps in the model parameters, with new models being accepted or rejected according to their probability computed as a Bayesian posterior probability. In this instance this probability is driven by a combination of $\chi^2$ and the prior probability from our spectroscopic constraints (see \citealt{parsons11b} and \citealt{pyrzas11} for more details of the MCMC fitting process).

For fitting the light curves of {\sdss}, we phase folded the data and kept the period fixed as one. We also kept the temperature of the white dwarf fixed at 17,900K (see Section~\ref{wdtemp}). For the secondary star we used quadratic limb darkening coefficients from \citet{claret11} for a $T_\mathrm{eff}=3000,\, \log{g}=5$ main sequence star. For the white dwarf we calculated quadratic limb darkening coefficients from a white dwarf model with $T_\mathrm{WD}=17,900$ and $\log{g}=7.53$ based on our spectroscopic fits, folded through the ULTRACAM $u'$, $g'$, $i'$ and SOFI $J$ filter profiles. For both stars we quote the coefficients $a$ and $b$ where $I(\mu)/I(1) = 1-a(1-\mu)-b(1-\mu)^2$, where $\mu$ is the cosine of the angle between the line of sight and the surface normal. We kept all limb darkening parameters fixed.

Table~\ref{sdss1212_lcfit} lists the best fit parameters from the light curves and their $1\sigma$ uncertainties, we also list the limb darkening coefficients used for each band. The results from all four bands are consistent. Figure~\ref{sdss1212_lcurves} shows the fits to the primary eclipses in each band and the residuals to the fits.

\subsection{{\gkvir}}

\begin{figure*}
  \begin{center}
    \includegraphics[width=0.9\textwidth]{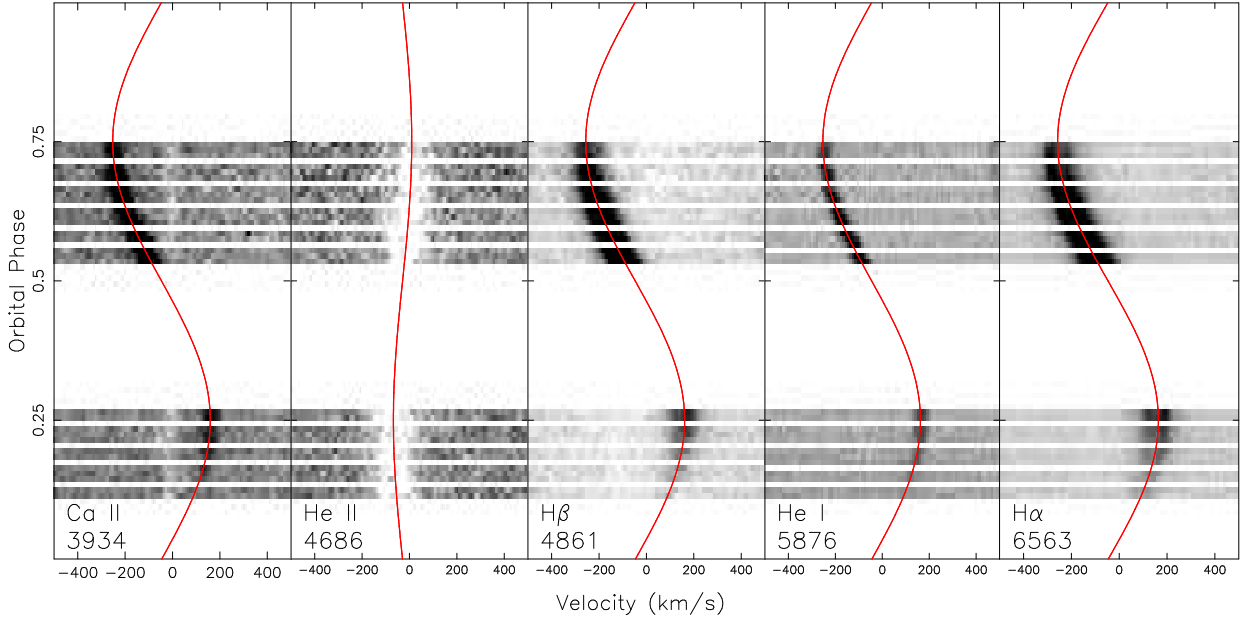}
    \vspace{4mm}
  \caption{Trailed spectra of several lines in {\gkvir}. The grey-scale runs from white (75 per cent of the continuum level) to black (125 per cent of the continuum level). The {\caii} 3934{\AA} line shows emission from the M star as well as weak interstellar absorption which shows no radial velocity variations. The {\heii} 4686{\AA} line originates from the white dwarf. The H$\beta$ and H$\alpha$ emission from the M star shows inverted cores and asymmetric profiles similar to those seen in NN\,Ser \citep{parsons10a}. The red lines (online version only) show the best fits to the lines.}
  \label{gkvir_trails}
  \end{center}
\end{figure*}

{\gkvir} (PG $1413+015$) was discovered by \citet{green78} from the Palomar-Green survey for ultraviolet-excess objects \citep{green86}. An eclipse was recorded during a subsequent spectroscopic observation. \citet{fulbright93} combined the photometry from \citet{green78} and high resolution spectroscopy to constrain the system parameters. They found that {\gkvir} contains a hot DAO white dwarf with a low-mass M3-5 main sequence companion in an $8^\mathrm{h}16^\mathrm{m}$ period. However, their lack of radial velocity information limited their analysis.

\subsubsection{Spectral features}
Figure~\ref{gkvir_spec} shows an average X-shooter spectrum of {\gkvir}. The hot white dwarf dominates the spectrum at wavelengths shorter than 1 micron. In the $J$ band the secondary star contributes roughly 50 per cent of the overall flux. Hydrogen Balmer absorption lines from the white dwarf are visible as well as narrow {\heii} 4686{\AA} absorption from the white dwarf, making it a DAO white dwarf, as mentioned by \citet{fulbright93}. Numerous emission lines originating from the heated face of the secondary star are seen throughout the spectrum. Na\,{\sc i} (8183{\AA}, 8195{\AA}) absorption originating from the secondary star is seen before phase 0.25 but decreases in strength towards phase 0.5 due to the increased ionisation and therefore cannot be used to measure the radial velocity amplitude of the secondary star. No other absorption features for the secondary star are visible. 

\subsubsection{Atmospheric parameters of the white dwarf}

We fitted the X-Shooter spectrum, averaged in the white dwarf rest-frame, to obtain an estimate of the effective temperature. As for {\sdss} (see Section~\ref{wdtemp}), we fitted the normalised Balmer lines, including H$\beta$ to H$\zeta$ and down-weighting the regions that are noticeably contaminated by emission lines. The best-fit gives a temperature of $55995\pm673$K, a surface gravity of $\log g=7.68\pm0.04$ and a distance of $550\pm20$\,pc. We note that the quoted errors are purely statistical, and that systematic effects are very likely affecting these results. The X-Shooter data were obtained near the quadrature phases ($\phi\simeq0.25$ and 0.75, in order to measure $K_\mathrm{WD}$ and $K_\mathrm{sec}$), which results in a significant contamination by emission from the strongly irradiated inner hemisphere of the secondary star. Varying the wavelength range around the lines that is down-weighted affects the resulting temperature by several 1000\,K. For comparison, we also fitted the lower-resolution SDSS spectrum of {\gkvir}, obtained near superior conjunction of the companion, and find $T_\mathrm{eff}=52258\pm3131$\,K, $\log g = 7.66\pm0.18$, and $d=509\pm61$\,pc.

\citet{fulbright93} analysed a blue spectrum of {\gkvir} obtained near the eclipse ($\phi\simeq0.02$), i.e. when the heated inner hemisphere of the companion contributes least to the observed flux. By fitting the H$\beta$-H$\zeta$ Balmer lines and the {\heii} 4686{\AA} line they determined $T_\mathrm{eff}=48800\pm1200$\,K and $\log g=7.70\pm0.11$. They also modelled the single available far-ultraviolet spectrum of {\gkvir}, and obtained $T_\mathrm=50000$\,K. A major limitation of this spectrum obtained with \textit{IUE} was that the photospheric Ly$\alpha$ line was nearly completely filled in by geocoronal emission.

We conclude that the effective temperature and distance of the white dwarf {\gkvir} remains somewhat uncertain, $T_\mathrm{eff}\simeq50000$\,K and $d\simeq500\pm50$\,pc. However, it is reassuring that all the spectroscopic measurements of the surface gravity are consistent with the value that is determined from the light curve fit (see Section~\ref{sec:discuss}).

Fitting the {\heii} 4686{\AA} absorption line from the white dwarf in {\gkvir} gives a helium abundance of $\log[\mathrm{He}/\mathrm{H}]=-2.8\pm0.3$ by numbers. Assuming that the secondary star transfers material of solar composition, the total accretion rate is $\dot M\simeq1.4\times10^9\mathrm{g\,s^{-1}}$ or $2.2\times10^{-17}\msy$, once again consistent with other pre-CVs.

\subsubsection{White dwarf radial velocity}
\begin{figure}
  \begin{center}
    \includegraphics[width=\columnwidth]{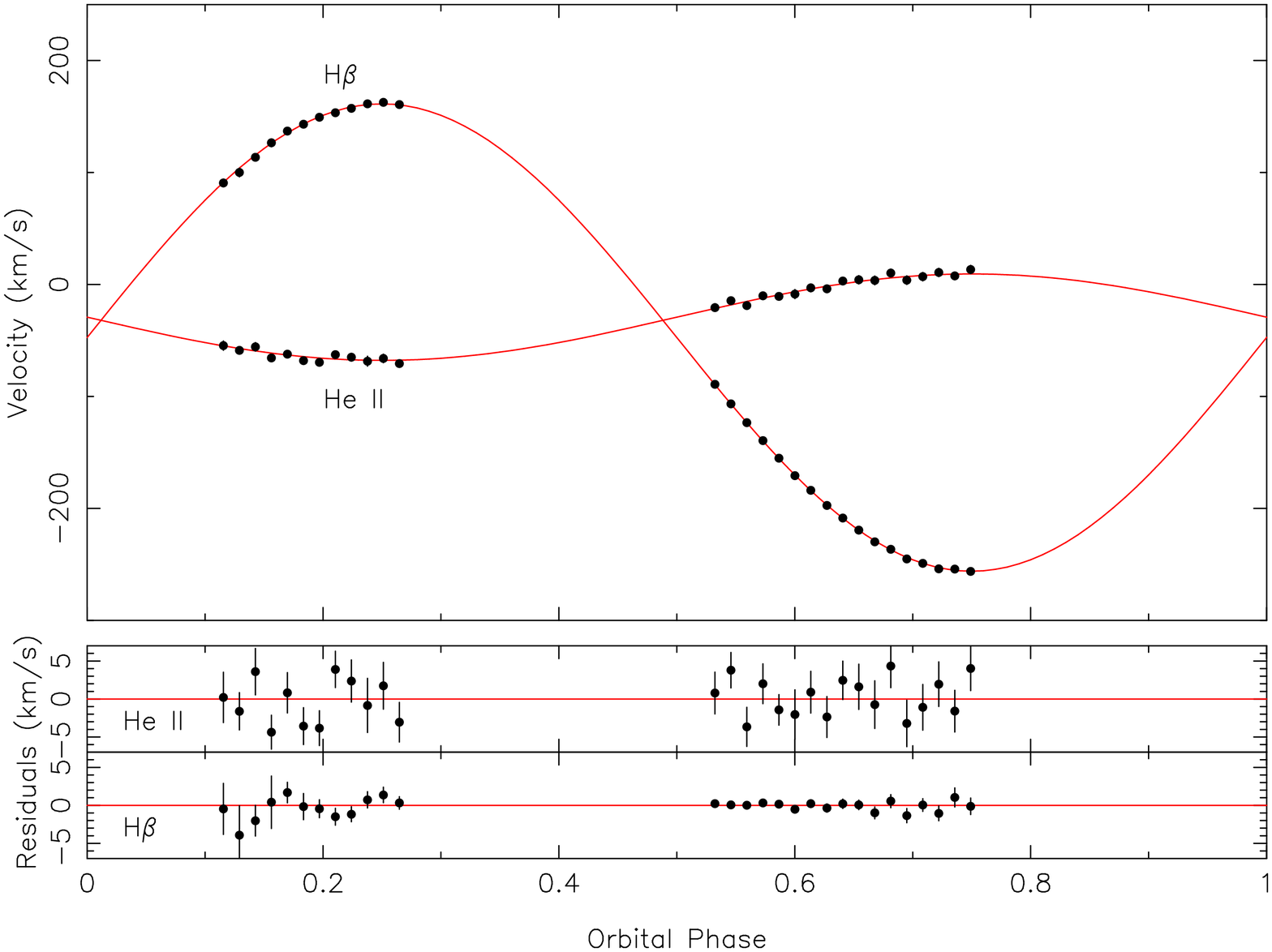}
    \vspace{1mm}
  \caption{Radial velocity fits to the {\heii} 4686{\AA} absorption from the white dwarf and H$\beta$ emission from the secondary star in {\gkvir} with residuals plotted below. The emission component does not track the centre of mass of the secondary star (see Section~\ref{kcorr}).}
  \label{gkvir_rvs}
  \end{center}
\end{figure}

Due to the high temperature of the white dwarf, the Balmer lines lack narrow cores and are therefore unsuitable for radial velocity work. We measured the radial velocity of the white dwarf in {\gkvir} from the {\heii} 4686{\AA} absorption line using the same technique as we used for {\sdss}.  Figure~\ref{gkvir_trails} shows a trail of the {\heii} line and Figure~\ref{gkvir_rvs} shows the fitted radial velocity curve, which gives a value of $K_\mathrm{WD} = 38.6\pm0.8\,\kms$ and a velocity offset of $\gamma_\mathrm{WD} = -27.2\pm0.7\,\kms$.

\begin{figure*}
  \begin{center}
    \includegraphics[width=0.98\columnwidth]{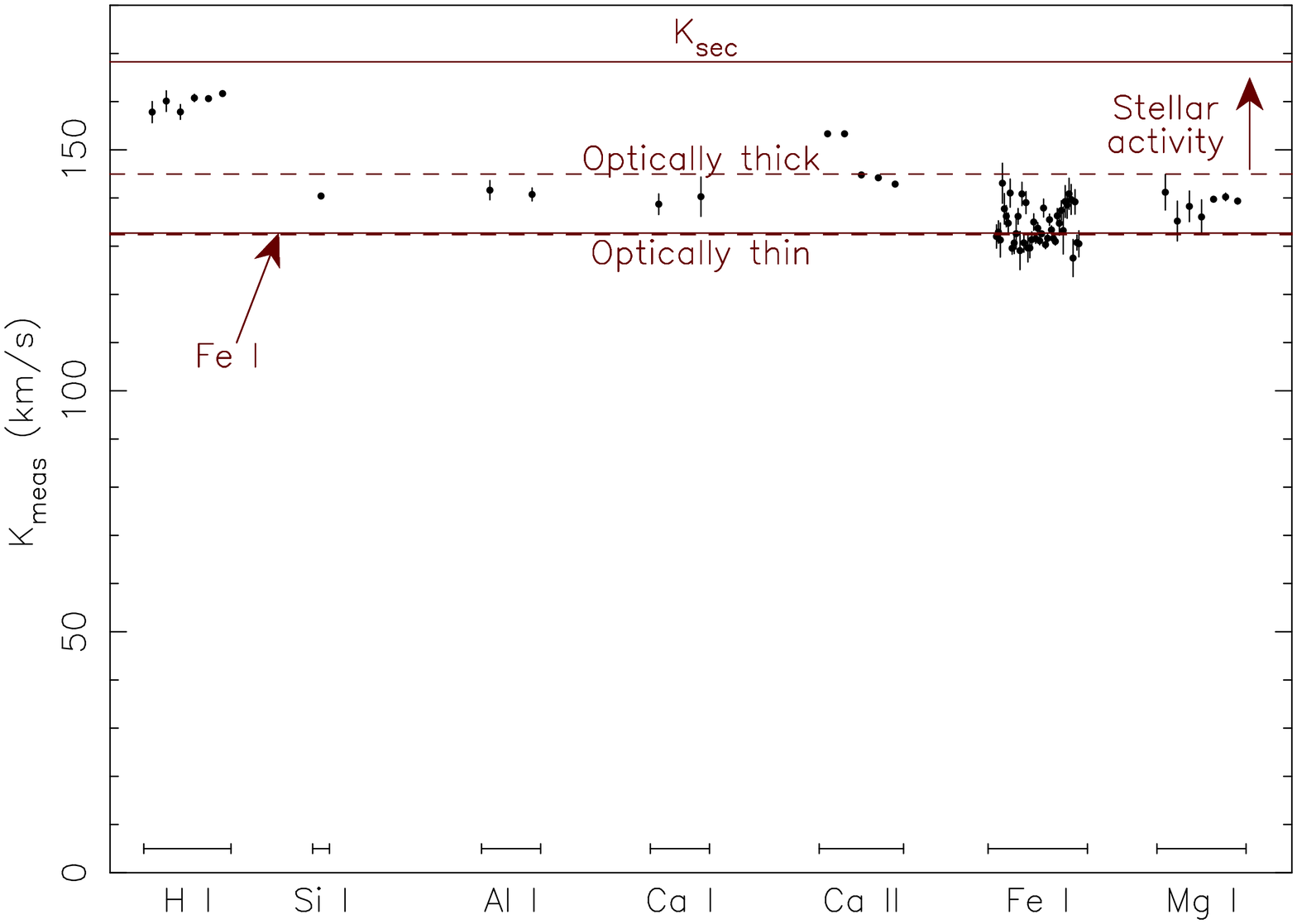}
    \hspace{2mm}
    \includegraphics[width=0.98\columnwidth]{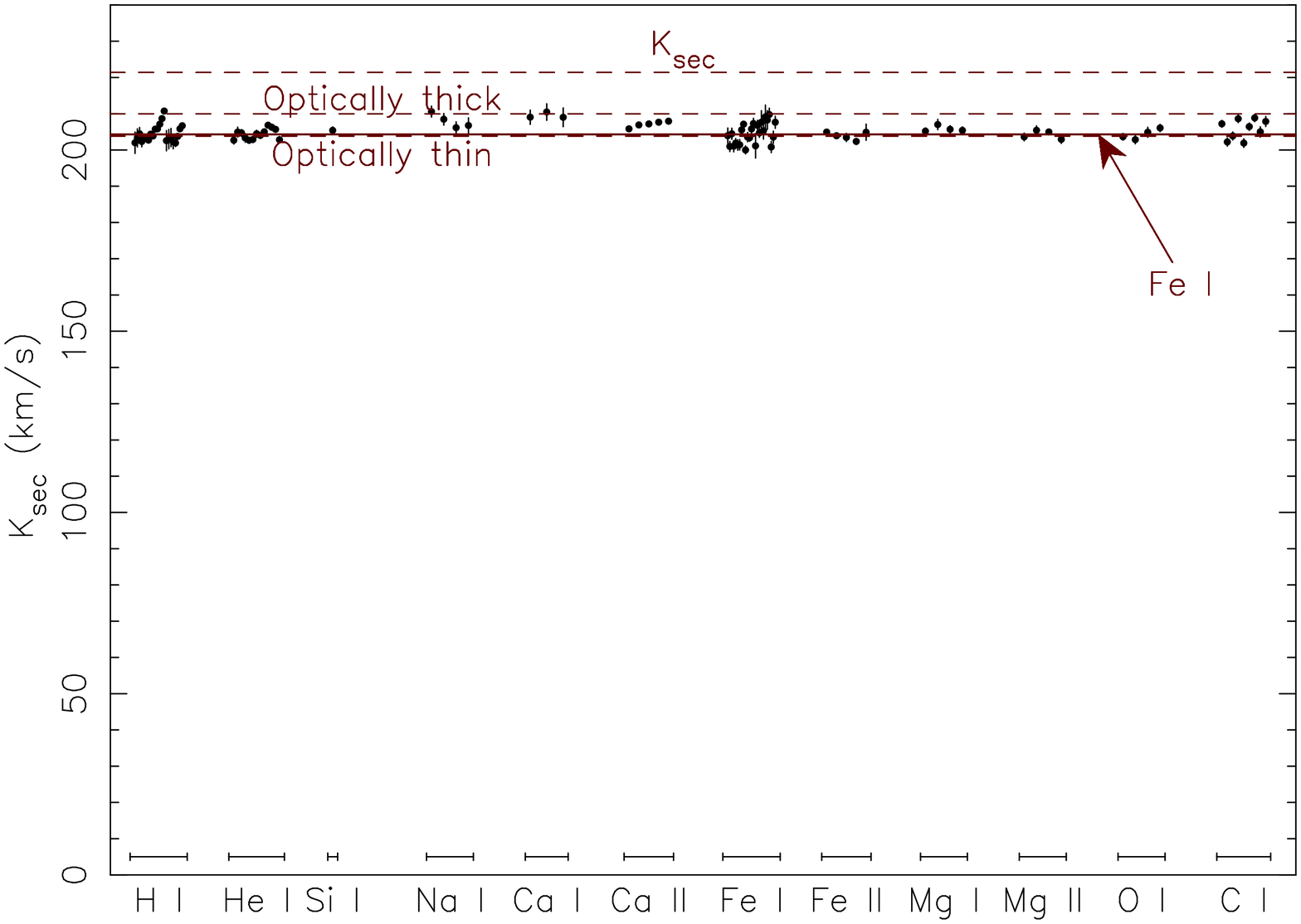}
  \caption{\emph{Left:} radial velocities of the emission lines from the secondary star in {\sdss}. The measured value of centre of mass radial velocity ($K_\mathrm{sec}$) is also shown. The two dashed lines denote the radial velocity that an optically thick and optically thin line would have. The solid line at $133\,\kms$ is a weighted average of the radial velocity amplitudes of the Fe\,{\sc i} lines, these appear to be optically thin. The radial velocity amplitudes of the hydrogen Balmer lines and the calcium H and K lines are far above the optically thick limit implying that these lines originate, at least in part, from stellar activity. \emph{Right:} a similar plot but for {\gkvir}. However, in this case we were unable to directly measure $K_\mathrm{sec}$ from the spectroscopy, but by assuming that the Fe\,{\sc i} lines are optically thin we get a consistent solution. The much tighter limits and smaller correction in {\gkvir} imply that the secondary star is quite small.}
  \label{kcorr_fig}
  \end{center}
\end{figure*}

\subsubsection{Emission lines}
A large number of emission lines are seen in the spectrum of {\gkvir} caused by the heating of the secondary star by the white dwarf. However, the secondary star in {\gkvir} receives a much larger irradiating flux to that in {\sdss} due to the higher temperature of the white dwarf, therefore a larger number of lines are present and from higher ionised states. Unfortunately the strength of the Na\,{\sc i} absorption doublet is too low after phase 0.25 and, combined with the lack of any other absorption features from the secondary star, means that we are unable to get a direct measurement of the radial velocity amplitude of the centre of mass of the secondary star. We determine the radial velocities and offsets of all of the emission lines identified using the same method used for {\sdss}, these are listed in Table~\ref{sec_emis_gkvir} in the Appendix. We use these results to determine a velocity offset of $\gamma_\mathrm{sec} = -47.35\pm0.05\,\kms$, giving a measured redshift for the white dwarf of $V_z=20.2\pm0.7\,\kms$.

\subsubsection{$K_\mathrm{sec}$ correction}
\label{kcorr}

As previously noted the emission lines in the X-shooter spectra of {\gkvir} cannot be used to directly measure $K_\mathrm{sec}$, needed for accurate mass determinations. We need to determine the deviation between the reprocessed light centre and the centre of mass for the secondary star. The radial velocity of the centre of mass ($K_\mathrm{sec}$) is related to that of the emission lines ($K_\mathrm{emis}$) by
\begin{eqnarray}
\label{eqn:kcorr}
K_\mathrm{sec} = \frac{K_\mathrm{emis}}{1 - f(1+q) R_\mathrm{sec}/a}
\end{eqnarray}
\citep{parsons11b}, where $f$ is a constant between 0 and 1 which depends upon the location of the centre of light. For $f=0$ the emission is spread uniformly across the entire surface of the secondary star and therefore the centre of light is the same as the centre of mass. For $f=1$ all of the flux is assumed to come from the point on the secondary star's surface closest to the white dwarf (the substellar point).

The centre of light for an emission line is related to the optical depth of the emission (\citealt{parsons10a}; \citealt{parsons11b}). Optically thick emission tends to be preferentially radiated perpendicular to the stellar surface, therefore at the quadrature phases, we will see the limb of the irradiated region more prominently (compared to the region of maximum irradiation) than we would otherwise. This will lead to a higher observed semi-amplitude and hence a smaller correction factor is needed. The reverse is true for optically thin lines where the emission is radiated equally in all directions, hence emission from the substellar point becomes enhanced at quadrature, leading to a low semi-amplitude and a larger correction factor. The correction factor for an optically thin line (assuming emissivity proportional to the incident flux) is $f=0.77$ and $f=0.5$ for optically thick emission \citep{parsons10a}. 

We can estimate the optical depths of the emission lines in {\gkvir} using our observation of {\sdss} and assuming a similar behaviour in the lines. We can determine the optical depth of the lines in {\sdss} because in this case we have a direct measurement of $K_\mathrm{sec}$. Therefore, we can reverse Eq (\ref{eqn:kcorr}) to determine the optical depths. The left hand panel of Figure~\ref{kcorr_fig} shows where the emission lines lie with respect to the optically thin lower limit and the optically thick upper limit. We find that the measured radial velocities of the Fe\,{\sc i} lines are consistent with them being optically thin. The majority of the emission lines in {\sdss} appear to lie somewhere between optically thick and optically thin with the exception of the hydrogen Balmer lines and the calcium H and K lines. For these lines the measured radial velocity is far higher than the optically thick limit, implying that the emission is more uniformly spread over the surface of the secondary star, this is most likely caused by stellar activity. This is consistent with the observations of several small flares during our photometry and implies that the secondary star in {\sdss} is an active star. 

From our analysis of the emission lines in {\sdss} we make the assumption that the Fe\,{\sc i} lines in {\gkvir} are also optically thin. We can then use their radial velocities to predict $K_\mathrm{sec}$ using Eq (\ref{eqn:kcorr}). We adopt a value for the radial velocity amplitude of an optically thin line of $K_\mathrm{Thin} = 204\pm2\,\kms$. We use this result as a prior constraint in our light curve fitting. The right hand panel of Figure~\ref{kcorr_fig} shows the radial velocity amplitudes of the emission lines in {\gkvir}, the dashed lines are based on the results of our light curve fitting (see section~\ref{gkvir_lcfitting}). The spread in the emission line radial velocities in {\gkvir} is relatively small compared to {\sdss} because of the small relative size of the secondary star ($R_\mathrm{sec}/a$).

\subsubsection{Eclipse times}
\label{ephem_gkvir}

\begin{table}
 \centering
  \caption{Eclipse times for {\gkvir}. (1) \citet{green78}, (2) \citet{parsons10b}, (3) this paper.}
  \label{gkvir_times}
  \begin{tabular}{@{}lcc@{}}
  \hline
Cycle & MJD(BTDB)         & Reference \\ 
No.   & (mid-eclipse)     &           \\
 \hline
 -67  & 42520.26747(1)    & (1) \\
 -32  & 42532.31905(2)    & (1) \\
 -29  & 42533.35204(9)    & (1) \\
   0  & 42543.33769(1)    & (1) \\
   3  & 42544.37068(1)    & (1) \\
 851  & 42836.36314(6)    & (1) \\
1966  & 43220.29202(12)   & (1) \\
2132  & 43277.45101(6)    & (1) \\
2896  & 43540.51972(12)   & (1) \\
28666 & 52413.9255716(9)  & (2) \\
29735 & 52782.0152272(9)  & (2) \\
29738 & 52783.0482185(7)  & (2) \\
30746 & 53130.1336878(27) & (2) \\
32706 & 53805.0221154(23) & (2) \\
32709 & 53806.0551129(12) & (2) \\
34054 & 54269.1800868(3)  & (2) \\
37069 & 55307.3375852(11) & (3) \\
\hline
\end{tabular}
\end{table}

\begin{figure}
  \begin{center}
    \includegraphics[width=\columnwidth]{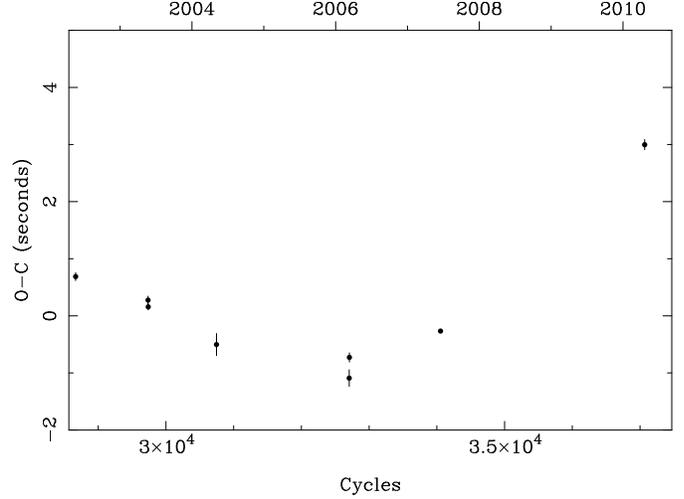}
    \vspace{3mm}
  \caption{Observed-Calculated (O-C) plot for the eclipse times of {\gkvir}. Our new point shows a departure from linearity in the eclipse times. The secondary star is just able to supply the energy required to drive this period change via Applegate's mechanism however, a third body in orbit around the system is also a possible explanation.}
  \label{gkvir_oc}
  \end{center}
\end{figure}

We recorded one new high precision eclipse time for {\gkvir} which is listed in Table~\ref{gkvir_times} along with all previous eclipse times. We update the ephemeris to 
\begin{eqnarray}
\mathrm{MJD(BTDB)} & = & 42543.337\,9121(33) \nonumber \\
                   &   & +\, 0.344\,330\,832\,742(99) E, \nonumber
\end{eqnarray}
which is consistent with previous studies. Figure~\ref{gkvir_oc} shows the difference between the observed eclipse time and the calculated eclipse time based on our new ephemeris. Our new eclipse time shows a clear deviation from linearity, although the magnitude of the period change is small compared to other systems (see \citealt{parsons10b}). The secondary star in {\gkvir} is able to drive this small period change (0.00124 seconds in $\sim8$ years) via Applegate's mechanism \citep{applegate92}, although this is only just the case if we used the modified version of Applegate's mechanism presented by \citet{brinkworth06} which takes account of the role of the inner part of the star in counterbalancing the outer shell. A third body in orbit around the system may also be the cause of the period change, as has recently been proposed for NN\,Ser \citep{beuermann10}, however, longer term monitoring is required in order to discover the true cause of this period change.

\subsubsection{Spectral type of the secondary star}

We detect the secondary star in {\gkvir} in the $r'$, $i'$ and $J$ band eclipses. We measure magnitudes of $r'=21.72\pm0.03$, $i'=19.98\pm0.01$ and $J=17.59\pm0.05$. The $r'-i'=1.74\pm0.03$ colour is consistent with a spectral type of M$4.5$ whilst the $i'-J=2.39\pm0.05$ is closer to a spectral type of M4 \citep{hawley02}. Using the distance from our spectroscopic fit ($550\pm20$pc) the secondary star has an absolute $J$ band magnitude of $M_J=8.9$, giving it a spectral type of M$4.5$ \citep{hawley02}, therefore we adopt a spectral type of M$4.5\pm0.5$.

\begin{figure*}
  \begin{center}
    \includegraphics[width=0.9\textwidth]{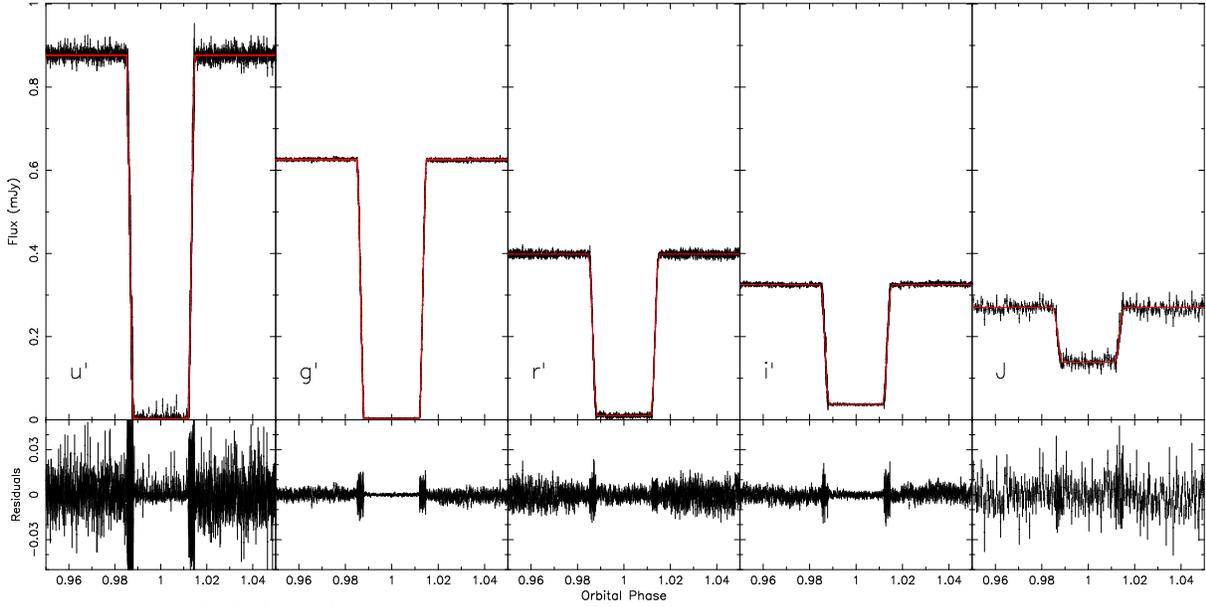}
    \vspace{2mm}
  \caption{ULTRACAM $u'$, $g'$, $r'$ and $i'$ and SOFI $J$ band primary eclipses of {\gkvir} with model fits and residuals. The light curves have been binned up by a factor of 5 although smaller bins were used on the ingress and egress features, which leads to the small spikes seen in the residuals at the ingress and egress.}
  \label{gkvir_lcurves}
  \end{center}
\end{figure*}

\begin{table*}
 \centering
  \caption{Parameters from Markov chain Monte Carlo minimisation for {\gkvir}, some fitted, some fixed a priori (those without quoted uncertainties). $a$ and $b$ are the quadratic limb darkening coefficients. $A$ is the fraction of the irradiating flux from the white dwarf absorbed by the secondary star, this was kept fixed for the $r'$ band since there is little coverage out-of-eclipse in this band.}
  \label{gkvir_lcfit}
  \begin{tabular}{@{}lccccc@{}}
  \hline
 Parameter & $u'$ & $g'$ & $r'$ & $i'$ & $J$ \\
 \hline 
$i$ (deg)          & $89.5\pm0.7$        & $89.5\pm0.6$       & $89.3\pm0.9$         & $89.2\pm0.9$        & $88.8\pm1.4$        \\
$r_\mathrm{WD}/a$    & $0.0093\pm0.0002$   & $0.0094\pm0.0002$  & $0.0093\pm 0.0003$   & $0.0094\pm0.0003$   & $0.0097\pm0.0006$   \\
$r_\mathrm{sec}/a$   & $0.085\pm0.002$     & $0.085\pm0.002$     & $0.086\pm0.002$     & $0.086\pm0.003$     & $0.088\pm0.006$     \\
$T_\mathrm{eff,sec}$ (K) & $3113\pm31$         & $3270\pm12$         & $3224\pm28$          & $3508\pm40$         & $4117\pm139$        \\
$a_\mathrm{WD}$      & $0.0769$            & $0.0594$            & $0.0505$             & $0.0446$            & $0.0257$            \\
$b_\mathrm{WD}$      & $0.1393$            & $0.1165$            & $0.0896$             & $0.0736$            & $0.0490$            \\
$a_\mathrm{sec}$     & $0.5866$            & $0.6720$            & $0.6364$             & $0.4193$            & $0.0254$            \\
$b_\mathrm{sec}$     & $0.2959$            & $0.2660$            & $0.2521$             & $0.4109$            & $0.4826$            \\
$A$                & $1.055\pm0.039$     & $0.551\pm0.006$     & $0.400$              & $0.263\pm0.009$     & $0.245\pm0.032$     \\
$K_\mathrm{sec}$    & $221.0\pm2.1$       & $221.2\pm2.1$       & $221.1\pm2.1$        & $221.3\pm2.1$       & $222.5\pm2.3$       \\
 \hline
\end{tabular}
\end{table*} 

\subsubsection{Modelling the light curves}
\label{gkvir_lcfitting}

The light curves of {\gkvir} show a deep eclipse of the white dwarf even in the $J$ band. {\gkvir} shows a small reflection effect out of eclipse, however no secondary eclipse is detected. We fit the light curves of {\gkvir} in the same way as those of {\sdss} (see Section~\ref{lcurvemodel}). However, since we lack a direct measurement of $K_\mathrm{sec}$, we use our constraint on $K_\mathrm{Thin}$ and Eq (\ref{eqn:kcorr}) as well as our $K_\mathrm{WD}$ and gravitational redshift measurements to constrain the inclination.

We phase binned the data, using smaller bins on the ingress and egress, and kept the period fixed as one when fitting the light curves. We also kept the temperature of the white dwarf fixed at 55,995K. For the secondary star we used quadratic limb darkening coefficients from \citet{claret11} for a $T_\mathrm{eff}=3000\mathrm{K},\, \log{g}=5$ main sequence star. For the white dwarf we calculated quadratic limb darkening coefficients from a white dwarf model with $T_\mathrm{WD}=55,995$ and $\log{g}=7.68$ based on our spectroscopic fits, folded through the ULTRACAM $u'$, $g'$, $r'$ and $i'$ and SOFI $J$ filter profiles. For both stars we quote the coefficients $a$ and $b$ where $I(\mu)/I(1) = 1-a(1-\mu)-b(1-\mu)^2$, where $\mu$ is the cosine of the angle between the line of sight and the surface normal, we kept all limb darkening parameters fixed. For our $r'$ band light curve we lack any out of eclipse information (barring that immediately before and after the eclipse) hence $A$, the fraction of irradiating flux absorbed by the secondary star, is unconstrained. Therefore, we fix this value at $0.4$; this parameter has no effect on the radii or inclination.

Figure~\ref{gkvir_lcurves} shows the fits to the light curves and the residuals and our final fitted parameters are listed in Table~\ref{gkvir_lcfit}. We also list the limb darkening coefficients used for each band.

\section{Discussion} \label{sec:discuss}

Our light curve fits combined with Eqs ({\ref{eqn:redshift1})-(\ref{eqn:redshift2}) yield direct measurements for the masses and radii of both components in {\sdss} and {\gkvir}. For our final values and uncertainties we combine the results of all our light curves, however, since they are all constrained by the same spectroscopic information the uncertainties in each light curve fit are not independent. Therefore we combine the results from each light curve in an optimal way by calculating the minimum possible error on each parameter purely from the spectroscopic constraints and then combining parameters allowing for the correlated and random noise components. Our final parameters for both systems are listed in Table~\ref{sys_para}. The secondary star's shape in both systems is slightly distorted due to the presence of the nearby white dwarf, therefore Table~\ref{sys_para} lists the radius of the secondary star in various directions. This is a very minor effect in {\gkvir} whilst the effect is somewhat larger for {\sdss}, though in both cases the variations are smaller than the uncertainty on the radius from the light curve fits. For our final discussions we adopt the volume-averaged radii.

\begin{table*}
 \centering
  \caption{System parameters. The surface gravities quoted are from the spectroscopic fit. They are consistent with the measured masses and radii. The accretion rate is that of the material from the wind of the secondary star onto the white dwarf}
  \label{sys_para}
  \begin{tabular}{@{}lll@{}}
  \hline
  Parameter         & {\sdss}                    & {\gkvir} \\
  \hline
  Period (days)     & 0.335\,871\,14(13)         & 0.344\,330\,832\,742(99) \\
  Inclination       & $85.7^{\circ}\pm0.5^{\circ}$  & $89.5^{\circ}\pm0.6^{\circ}$ \\
  Binary separation & $1.815\pm0.003$ R$_{\sun}$  & $1.82\pm0.01$ R$_{\sun}$ \\
  Mass ratio        & $0.620\pm0.001$            & $0.174\pm0.004$ \\
  WD mass           & $0.439\pm0.002$ M$_{\sun}$  & $0.564\pm0.014$ M$_{\sun}$ \\
  Sec mass          & $0.273\pm0.002$ M$_{\sun}$  & $0.116\pm0.003$ M$_{\sun}$ \\
  WD radius         & $0.0168\pm0.0003$ R$_{\sun}$& $0.0170\pm0.0004$ R$_{\sun}$ \\
  Sec radius polar           & $0.304\pm0.007$ R$_{\sun}$ & $0.154\pm0.003$ R$_{\sun}$ \\
  Sec radius sub-stellar     & $0.310\pm0.007$ R$_{\sun}$ & $0.156\pm0.003$ R$_{\sun}$ \\
  Sec radius backside        & $0.309\pm0.007$ R$_{\sun}$ & $0.156\pm0.003$ R$_{\sun}$ \\
  Sec radius side            & $0.306\pm0.007$ R$_{\sun}$ & $0.155\pm0.003$ R$_{\sun}$ \\
  Sec radius volume-averaged & $0.306\pm0.007$ R$_{\sun}$ & $0.155\pm0.003$ R$_{\sun}$ \\
  WD $\log{g}$      & $7.51\pm0.01$              & $7.68\pm0.04$ \\
  WD temperature    & $17,707\pm35$K             & $55,995\pm673$K \\
  $K_\mathrm{WD}$    & $104.4\pm0.5\,\kms$        & $38.6\pm0.8\,\kms$  \\
  $K_\mathrm{sec}$   & $168.3\pm0.3\,\kms$        & $221.6\pm2.0\,\kms$ \\
  $V_{z,\mathrm{WD}}$  & $16.1\pm0.3\,\kms$         & $20.2\pm0.7\,\kms$ \\
  Sec spectral type & M4                         & M$4.5\pm0.5$ \\
  Distance          & $228\pm5\,$pc              & $550\pm20\,$pc \\
  Accretion rate    & $6.4\times10^{-17}\msy$     & $2.2\times10^{-17}\msy$ \\
 \hline
\end{tabular}
\end{table*}

\begin{figure}
  \begin{center}
    \includegraphics[width=\columnwidth]{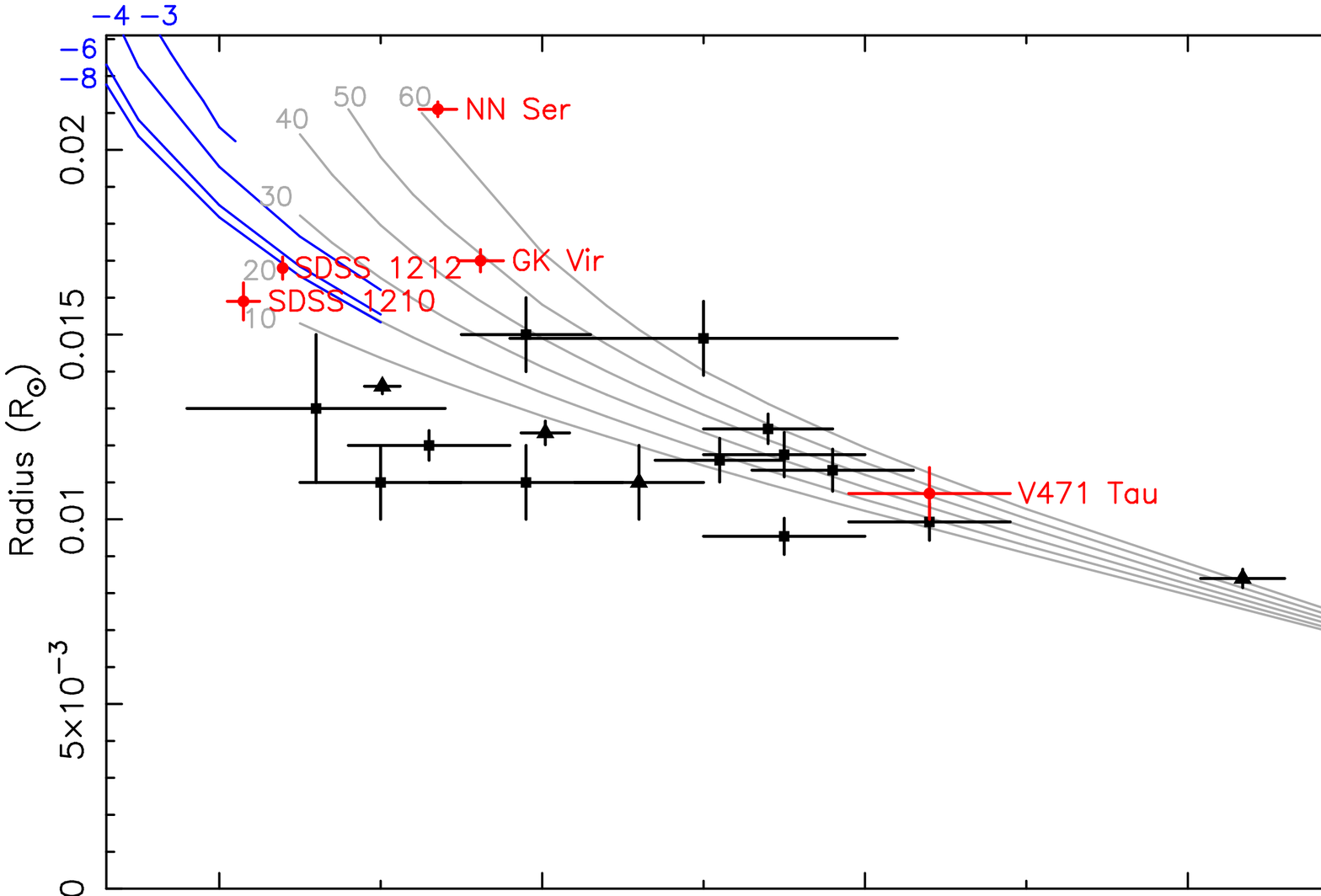}
    \vspace{8mm}
  \caption{Mass-radius plot for white dwarfs. Black points are from \citet{provencal98}, \citet{provencal02} and \citet{casewell09}. The square points are common proper-motion systems, the triangular points are visual binaries. White dwarfs measurements from PCEBs are shown in red (online version only) from \citep{obrien01}, \citet{parsons10a}, \citet{pyrzas11} and this work. The grey lines are CO core models with the temperatures labelled in units of $10^3$K and with hydrogen layer thicknesses of M$_\mathrm{H}/\mathrm{M}_\mathrm{WD} = 10^{-4}$ from \citet{benvenuto99}. The mass and radius of the white dwarf in {\gkvir} agree well with the $50,000$K model. The blue lines (online version only) are He core models with a temperature of $18,000$K and varying hydrogen layer thicknesses labelled by the exponent of the hydrogen layer fraction, from \citet{panei07}. The mass and radius of the white dwarf in {\sdss} is consistent with the CO core models for its temperature but we can rule out a CO core on evolutionary grounds. Therefore the white dwarf in {\sdss} is consistent with the He core models only if it has a very thin hydrogen envelope (M$_\mathrm{H}/\mathrm{M}_\mathrm{WD} \leq 10^{-6}$).}
  \label{WDMR}
  \end{center}
\end{figure}

Figure~\ref{WDMR} shows the mass-radius plot for white dwarfs. The position of the white dwarfs in {\sdss} and {\gkvir} are shown as well as other accurate white dwarf mass-radius measurements. The measured mass and radius of the white dwarf in {\gkvir} are consistent with a carbon-oxygen (CO) core white dwarf of the same temperature, with a thick hydrogen envelope (M$_\mathrm{H}/\mathrm{M}_\mathrm{WD} = 10^{-4}$). The measured mass and radius of the white dwarf in {\sdss} are also consistent with a CO core white dwarf with the same temperature and a thick hydrogen envelope (M$_\mathrm{H}/\mathrm{M}_\mathrm{WD} = 10^{-4}$). However, although it is possible to create CO core white dwarfs with masses $<0.5$ M$_{\sun}$ via considerable mass loss along the red giant phase (\citealt{prada09}; \citealt{willems04}; \citealt{han00}), doing so in a binary system requires a large initial mass ratio and results in a widening of the orbital separation, hence we would not expect a CO core white dwarf with a mass $<0.5$ M$_{\sun}$ in a close binary system. Therefore the white dwarf in {\sdss} must have a He core, as noted by \citet{shen09}.

Several He core mass-radius relations are shown in Figure~\ref{WDMR} for a white dwarf with a temperature of $18,000$K and varying hydrogen layer thicknesses. The white dwarf in {\sdss} is consistent with these relations only if it has a very thin hydrogen envelope (M$_\mathrm{H}/\mathrm{M}_\mathrm{WD} \leq 10^{-6}$). Figure~\ref{sdss1212_MR} shows a zoomed in version of the mass radius plot for the white dwarf in {\sdss} as well as the same He core models as Figure~\ref{WDMR}. The black line shows the range of possible masses and radii that the white dwarf in {\sdss} could have based on the radial velocities and the primary eclipse shape (i.e. no inclination constraints). The numbers plotted along this line are what the measured gravitational redshift of the white dwarf would need to be in order to give that mass and radius. Previous studies have found that the spectroscopic gravitational redshift measurements are usually slightly inconsistent with mass-radius measurements via other methods (\citealt{pyrzas11}; \citealt{parsons10a}; \citealt{maxted07}) meaning that our inclination constraints may be slightly incorrect. However, Figure~\ref{sdss1212_MR} shows that it is not possible for the white dwarf to have a thick hydrogen envelope (M$_\mathrm{H}/\mathrm{M}_\mathrm{WD} > 2 \times 10^{-4}$) even if the inclination is $90^{\circ}$. Current evolutionary models are unable to create He core white dwarfs with such thin hydrogen envelopes, the thinnest envelopes are of the order of M$_\mathrm{H}/\mathrm{M}_\mathrm{WD} \sim 3 \times 10^{-4}$ (\citealt{althaus09}; \citealt{sarna00}; \citealt{driebe98}) meaning that, assuming our measured $V_z$ value is accurate, either {\sdss} has had a very unusual evolutionary history or that current evolutionary models of He core white dwarfs are incomplete and overestimate their size. Additional He core white dwarf mass-radius measurements should show if this is the case. 

\begin{figure}
  \begin{center}
    \includegraphics[width=\columnwidth]{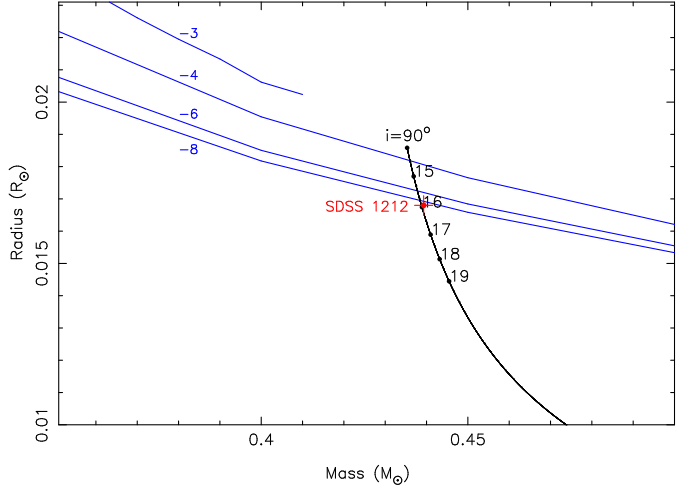}
    \vspace{10mm}
  \caption{Mass-radius plot for the white dwarf in {\sdss}. The blue lines (online version only) are He core models with a temperature of $18,000$K and varying hydrogen layer thicknesses labelled by the exponent of the hydrogen layer fraction, from \citet{panei07}. The black line shows the possible mass and radius range without any constraint on the inclination with $90^\circ$ at the top. The numbers along this line correspond to the gravitational redshift that the white dwarf would have at that inclination (in $\kms$). Our measured redshift ($16.1\pm0.3\,\kms$) means that the white dwarf is consistent with models which have a thin hydrogen layer. However, even without any inclination constraints it is not possible for the white dwarf to have a thick (M$_\mathrm{H}/\mathrm{M}_\mathrm{WD} > 2 \times 10^{-4}$) hydrogen layer.}
  \label{sdss1212_MR}
  \end{center}
  \vspace{10mm}
\end{figure}

Figure~\ref{MDMR} show the mass-radius plot for low mass stars. The masses and radii of the secondary stars in {\sdss} and {\gkvir} are marked as well as other precise measurements. The mass and radius of the secondary star in {\sdss} show that it is over-inflated for its mass compared with evolutionary models by $\sim12$ per cent. This discrepancy can be reduced to $\sim9$ per cent if the secondary star is active. In this case the radius may be overestimated due to the effects of polar spots, it also increases to compensate for loss of radiative efficiency due to starspots and due to a strong magnetic field caused by rapid rotation \citep{morales10,chabrier07}. We have evidence from the emission lines and light curves that the secondary star in {\sdss} is indeed active, therefore these effects can explain some of the discrepancy however, the star remains oversized. The secondary star in {\gkvir} is also oversized by $\sim9$ per cent, this drops to $\sim6$ per cent if it is active, though we have no evidence of activity from this star.

Taking the effects of rotational and tidal perturbations fully into account only causes an increase in the radii of 0.1 per cent for {\gkvir} and 0.4 per cent for {\sdss} \citep{sirotkin09} which is not enough to explain the discrepancy.

Irradiation by the white dwarf can cause the secondary star to become inflated by effectively blocking the energy outflow through the surface layers \citep{ritter00}. For NN\,Ser correcting this effect brought the mass and radius measurements into agreement with evolutionary models \citep{parsons10a}. For the secondary star in {\gkvir} we find that irradiation increases its radius by 5.6\% \citep{ritter00,hameury97}, enough to bring it into agreement with evolutionary models. However, for the far less irradiated secondary star in {\sdss} we find an increase of only 0.4\% meaning that it is still overinflated. The corrected radii are shown in Figure~\ref{MDMR}

\begin{figure}
  \begin{center}
    \includegraphics[width=\columnwidth]{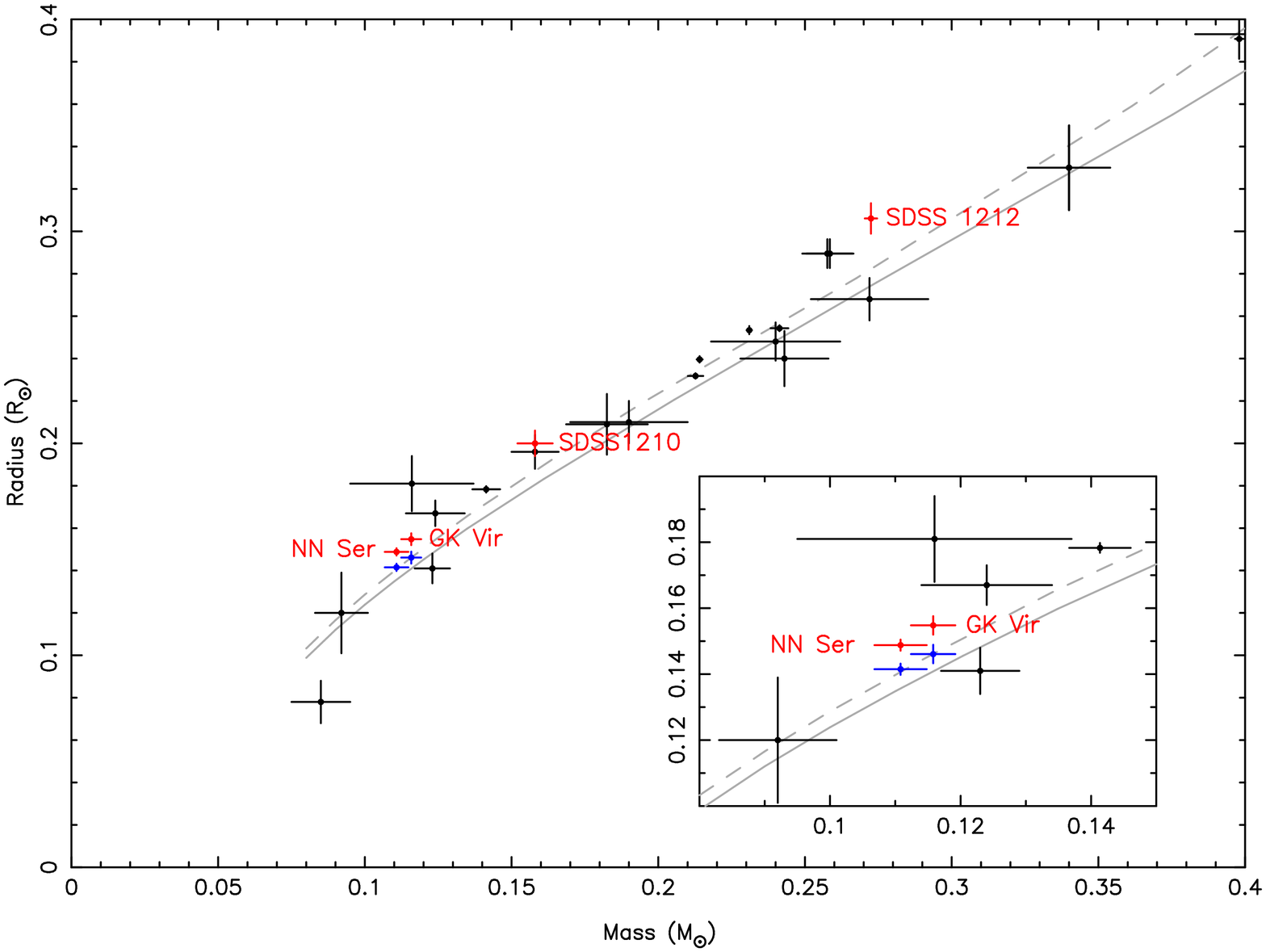}
  \caption{Mass-radius plot for low-mass stars. Black points are high-precision measurements taken from \citet{knigge11}, \citet{carter11} and \citet{ofir11}. The solid line is the 4.0-Gyr isochrone from \citet{baraffe98} whilst the dashed line is a 4.0-Gyr model from \citet{morales10} which includes the effects of magnetic activity. Low-mass stars from PCEBs are shown in red (online version only) from \citet{parsons10a}, \citet{pyrzas11} and this work. The blue points (online version only) show the radii of these stars after correcting for irradiation (negligible for SDSS\,J1210+3347 and {\sdss}). The secondary star in {\gkvir} is consistent with model predictions after correcting for irradiation, however the secondary star in {\sdss} is still oversized.}
  \label{MDMR}
  \end{center}
\end{figure}

\section{Conclusions}

We have used a combination of ULTRACAM and SOFI photometry and X-shooter spectroscopy to measure precise masses and radii for both components of the eclipsing PCEBs {\sdss} and {\gkvir}. In both cases we use measurements of the gravitational redshift of the white dwarf to constrain the orbital inclination. We were able to directly measure the radial velocity amplitudes of both stars in {\sdss} but only the white dwarf in {\gkvir}. However, we use our results from {\sdss} and information from the emission lines to determine the correction factor to apply to the emission line radial velocities in {\gkvir} to measure the centre of mass radial velocity of the secondary star. The mass and radius of the $0.564$ M$_{\sun}$ white dwarf in {\gkvir} are consistent with evolutionary models for a CO core white dwarf with a thick hydrogen envelope. The mass and radius of the white dwarf in {\sdss} place it in the crossover region between a CO core white dwarf and a He core white dwarf but we can exclude a CO core on evolutionary grounds. This means that the white dwarf in {\sdss} is the first He core white dwarf with precise mass-radius measurements however, it is under-inflated for its temperature unless it has a very thin (M$_\mathrm{H}/\mathrm{M}_\mathrm{WD} \leq 10^{-6}$) hydrogen envelope, which evolutionary models are unable to produce. The mass and radius of the secondary star in {\gkvir} are consistent with evolutionary models after correcting for the effects of irradiation by the white dwarf, however, the secondary star in {\sdss} is overinflated compared to theoretical predictions even after the effects of activity, rapid rotation and irradiation are taken into account.

\section*{Acknowledgements}
We thank the anonymous referee for their useful comments and suggestions. ULTRACAM, TRM, BTG, CMC, VSD and SPL are supported by the Science and Technology Facilities Council (STFC). SPL also acknowledges the support of an RCUK Fellowship. ARM acknowledges financial support from FONDECYT in the form of grant number 3110049. MRS thanks for support from FONDECYT (1100782). The results presented in this paper are based on observations collected at the European Southern Observatory under programme IDs 085.D-0541 and 087.D-0046. We used the National Institute of Standards and Technology (NIST) Atomic Spectra Database (version 4.0.1). 

\bibliographystyle{mn2e}
\bibliography{eclipsers}

\appendix

\section{Emission lines}

\begin{table*}
 \centering
  \caption{Secondary star emission lines in {\sdss}. After measuring the radial velocity of a line, the spectra were shifted to remove the motion and averaged. The Equivalent widths were measured from this averaged spectrum.}
  \label{sec_emis_1212}
  \begin{tabular}{@{}lccc@{}}
\hline
Line      & $\gamma_\mathrm{sec}$ & $K_\mathrm{meas}$ & Equivalent   \\
          & (km/s)          & (km/s)          & Width (m\AA) \\
\hline
Mg\,{\sc i}  3829.355 & $18.31\pm2.52$ & $141.19\pm3.68$ & $72\pm3$ \\ 
Mg\,{\sc i}  3832.299 & $22.90\pm3.20$ & $135.22\pm4.12$ & $60\pm3$ \\ 
Mg\,{\sc i}  3838.292 & $17.09\pm2.25$ & $138.27\pm3.17$ & $67\pm3$ \\ 
Fe\,{\sc i}  3878.573 & $18.74\pm1.89$ & $132.01\pm2.41$ & $83\pm2$ \\ 
H8           3889.055 & $17.47\pm1.73$ & $157.82\pm2.18$ & $74\pm3$ \\ 
Si\,{\sc i}  3905.523 & $22.79\pm0.33$ & $140.42\pm0.43$ & $248\pm2$ \\ 
Fe\,{\sc i}  3922.911 & $20.92\pm1.86$ & $132.93\pm2.32$ & $41\pm2$ \\ 
Fe\,{\sc i}  3928.083 & $15.52\pm2.49$ & $131.28\pm3.52$ & $37\pm2$ \\ 
Ca\,{\sc ii} 3933.663 & $22.31\pm0.38$ & $153.32\pm0.48$ & $386\pm3$ \\ 
Al\,{\sc i}  3944.006 & $25.21\pm1.41$ & $141.63\pm1.99$ & $48\pm2$ \\ 
Al\,{\sc i}  3961.520 & $19.95\pm1.01$ & $140.74\pm1.35$ & $72\pm2$ \\ 
Ca\,{\sc ii} 3968.469 & $21.05\pm0.53$ & $153.32\pm0.48$ & $349\pm3$ \\ 
H$\epsilon$  3970.074 & $17.03\pm2.88$ & $160.10\pm2.13$ & $84\pm2$ \\ 
H$\delta$    4101.735 & $21.62\pm1.23$ & $157.86\pm1.53$ & $262\pm4$ \\ 
Fe\,{\sc i}  4143.868 & $22.17\pm2.91$ & $143.06\pm4.15$ & $31\pm2$ \\ 
Fe\,{\sc i}  4216.184 & $20.09\pm2.30$ & $137.76\pm3.16$ & $27\pm2$ \\ 
Ca\,{\sc i}  4226.728 & $22.94\pm1.59$ & $138.71\pm2.13$ & $91\pm2$ \\ 
H$\gamma$    4340.465 & $17.31\pm0.60$ & $160.76\pm0.76$ & $320\pm3$ \\ 
Fe\,{\sc i}  4375.986 & $21.88\pm1.73$ & $136.22\pm2.20$ & $54\pm2$ \\ 
Fe\,{\sc i}  4427.310 & $19.64\pm1.57$ & $134.77\pm2.09$ & $49\pm2$ \\ 
Fe\,{\sc i}  4461.653 & $20.38\pm2.23$ & $141.02\pm2.91$ & $47\pm2$ \\ 
Fe\,{\sc i}  4482.170 & $20.86\pm1.09$ & $129.57\pm1.22$ & $36\pm2$ \\ 
Mg\,{\sc i}  4571.096 & $16.38\pm2.98$ & $136.07\pm3.55$ & $23\pm2$ \\ 
H$\beta$     4861.327 & $19.97\pm0.32$ & $160.61\pm0.40$ & $675\pm4$ \\ 
Fe\,{\sc i}  4924.298 & $19.42\pm1.46$ & $130.66\pm2.18$ & $65\pm2$ \\ 
Fe\,{\sc i}  4939.686 & $18.22\pm2.34$ & $132.61\pm3.18$ & $29\pm2$ \\ 
Fe\,{\sc i}  4957.597 & $19.55\pm1.07$ & $136.19\pm1.63$ & $48\pm1$ \\ 
Fe\,{\sc i}  5006.119 & $21.48\pm2.64$ & $129.10\pm3.95$ & $32\pm2$ \\ 
Fe\,{\sc i}  5012.068 & $15.55\pm1.77$ & $140.84\pm2.42$ & $46\pm2$ \\ 
Fe\,{\sc i}  5041.447 & $20.98\pm1.45$ & $130.69\pm1.91$ & $91\pm2$ \\ 
Fe\,{\sc i}  5051.664 & $18.19\pm1.64$ & $139.04\pm2.28$ & $41\pm2$ \\ 
Fe\,{\sc i}  5079.740 & $23.25\pm2.15$ & $129.63\pm2.89$ & $48\pm2$ \\ 
Fe\,{\sc i}  5083.338 & $19.40\pm1.56$ & $129.62\pm2.01$ & $30\pm2$ \\ 
Fe\,{\sc i}  5107.447 & $19.03\pm1.19$ & $131.36\pm1.63$ & $82\pm2$ \\ 
Fe\,{\sc i}  5110.413 & $17.30\pm1.29$ & $135.01\pm1.71$ & $87\pm2$ \\ 
Mg\,{\sc i}  5167.322 & $17.26\pm0.45$ & $139.73\pm0.59$ & $150\pm2$ \\ 
Fe\,{\sc i}  5168.898 & $19.56\pm0.75$ & $131.56\pm0.96$ & $78\pm2$ \\ 
Mg\,{\sc i}  5172.684 & $20.00\pm0.56$ & $140.22\pm0.78$ & $254\pm3$ \\ 
Mg\,{\sc i}  5183.604 & $17.46\pm0.46$ & $139.37\pm0.62$ & $209\pm2$ \\ 
Fe\,{\sc i}  5227.189 & $21.54\pm0.85$ & $133.72\pm1.16$ & $76\pm2$ \\ 
Fe\,{\sc i}  5269.537 & $22.13\pm0.61$ & $131.14\pm0.85$ & $173\pm2$ \\ 
Fe\,{\sc i}  5328.038 & $21.83\pm0.42$ & $132.58\pm0.58$ & $160\pm2$ \\ 
Fe\,{\sc i}  5341.234 & $22.18\pm1.35$ & $137.92\pm1.86$ & $83\pm2$ \\ 
Fe\,{\sc i}  5371.489 & $15.66\pm0.60$ & $130.31\pm0.82$ & $101\pm2$ \\ 
Fe\,{\sc i}  5397.128 & $17.84\pm0.63$ & $131.63\pm0.87$ & $103\pm2$ \\ 
Fe\,{\sc i}  5405.774 & $20.25\pm0.85$ & $135.49\pm1.16$ & $75\pm2$ \\ 
Fe\,{\sc i}  5429.695 & $20.10\pm0.60$ & $133.37\pm0.84$ & $97\pm2$ \\ 
Fe\,{\sc i}  5446.871 & $19.61\pm0.56$ & $131.62\pm0.74$ & $152\pm2$ \\ 
Fe\,{\sc i}  5455.609 & $20.32\pm0.56$ & $130.99\pm0.76$ & $121\pm2$ \\ 
Fe\,{\sc i}  5497.516 & $21.74\pm1.09$ & $136.31\pm1.49$ & $85\pm2$ \\ 
Fe\,{\sc i}  5501.465 & $18.64\pm1.78$ & $134.74\pm2.46$ & $52\pm3$ \\ 
Fe\,{\sc i}  5506.779 & $18.37\pm1.57$ & $137.41\pm2.02$ & $55\pm3$ \\ 
Na\,{\sc i}  5889.950 & $20.63\pm0.84$ & $144.01\pm1.13$ & $337\pm7$ \\ 
Na\,{\sc i}  5895.924 & $20.69\pm1.02$ & $144.35\pm1.42$ & $204\pm7$ \\ 
Fe\,{\sc i}  6136.994 & $21.08\pm3.63$ & $133.25\pm4.85$ & $149\pm7$ \\ 
Ca\,{\sc i}  6162.170 & $19.44\pm2.74$ & $140.27\pm4.07$ & $65\pm6$ \\ 
Fe\,{\sc i}  6191.588 & $20.43\pm2.43$ & $139.27\pm3.30$ & $87\pm7$ \\ 
Fe\,{\sc i}  6230.723 & $20.26\pm2.03$ & $138.54\pm2.75$ & $92\pm6$ \\ 
Fe\,{\sc i}  6252.555 & $16.48\pm2.40$ & $140.87\pm3.22$ & $52\pm5$ \\ 
Fe\,{\sc i}  6393.601 & $21.92\pm2.25$ & $139.69\pm3.10$ & $65\pm6$ \\ 
Fe\,{\sc i}  6400.001 & $21.00\pm2.89$ & $127.53\pm3.83$ & $89\pm7$ \\ 
Fe\,{\sc i}  6430.846 & $17.43\pm1.97$ & $139.20\pm2.53$ & $84\pm6$ \\ 
Fe\,{\sc i}  6494.980 & $18.62\pm1.14$ & $130.75\pm1.55$ & $106\pm5$ \\ 
\end{tabular}
\end{table*}

\begin{table*}
\centering
\begin{tabular}{@{}lccc@{}}
H$\alpha$    6562.760 & $18.91\pm0.16$ & $161.66\pm0.20$ & $2894\pm9$ \\ 
Fe\,{\sc i}  6677.987 & $23.21\pm2.15$ & $130.50\pm2.70$ & $71\pm5$ \\ 
Ca\,{\sc ii} 8498.020 & $20.35\pm0.13$ & $144.80\pm0.18$ & $887\pm4$ \\ 
Ca\,{\sc ii} 8542.090 & $20.13\pm0.14$ & $144.19\pm0.20$ & $847\pm4$ \\ 
Ca\,{\sc ii} 8662.140 & $19.87\pm0.17$ & $142.89\pm0.23$ & $671\pm4$ \\
\hline
\end{tabular}
\end{table*}

\begin{table*}
 \centering
  \caption{Secondary star emission lines in {\gkvir}. P is the hydrogen Paschen series. After measuring the radial velocity of a line, the spectra were shifted to remove the motion and averaged. The Equivalent widths were measured from this averaged spectrum.}
  \label{sec_emis_gkvir}
  \begin{tabular}{@{}lccc@{}}
\hline
Line      & $\gamma_\mathrm{sec}$ & $K_\mathrm{meas}$ & Equivalent   \\
          & (km/s)          & (km/s)          & Width (m\AA) \\
\hline
H16  3703.853         & $-45.55\pm2.07$ & $200.01\pm2.93$ & $68\pm3$ \\ 
H15  3711.971         & $-48.49\pm1.83$ & $203.33\pm2.49$ & $81\pm2$ \\ 
Fe\,{\sc i} 3719.935  & $-50.25\pm1.55$ & $197.94\pm2.10$ & $29\pm2$ \\ 
H14 3721.948          & $-46.01\pm1.21$ & $204.48\pm1.71$ & $107\pm2$ \\ 
H13 3734.372          & $-49.96\pm1.01$ & $199.31\pm1.42$ & $183\pm2$ \\ 
Fe\,{\sc i}  3737.131 & $-46.50\pm1.00$ & $199.97\pm1.40$ & $53\pm2$ \\ 
H12 3750.152          & $-46.68\pm0.73$ & $203.38\pm1.04$ & $197\pm2$ \\ 
H11 3770.634          & $-47.06\pm0.49$ & $203.61\pm0.70$ & $239\pm2$ \\ 
H10 3797.910          & $-46.32\pm0.39$ & $202.72\pm0.55$ & $311\pm2$ \\ 
He\,{\sc i}  3819.761 & $-48.22\pm0.73$ & $201.68\pm1.05$ & $69\pm2$ \\ 
H9  3835.397          & $-48.00\pm0.32$ & $204.37\pm0.46$ & $331\pm2$ \\ 
Fe\,{\sc i}  3856.371 & $-46.33\pm1.11$ & $204.48\pm1.52$ & $41\pm2$ \\ 
H8  3889.055          & $-47.28\pm0.21$ & $203.86\pm0.30$ & $457\pm2$ \\ 
Si\,{\sc i}  3905.523 & $-45.96\pm0.69$ & $205.38\pm0.97$ & $58\pm2$ \\ 
Ca\,{\sc ii} 3933.663 & $-46.67\pm0.28$ & $205.82\pm0.39$ & $208\pm2$ \\ 
He\,{\sc i}  3964.727 & $-48.70\pm0.93$ & $198.93\pm1.36$ & $44\pm2$ \\ 
Ca\,{\sc ii} 3968.469 & $-46.78\pm0.31$ & $206.90\pm0.40$ & $158\pm2$ \\ 
H$\epsilon$  3970.074 & $-47.53\pm0.20$ & $205.75\pm0.27$ & $529\pm2$ \\ 
He\,{\sc i}  4026.189 & $-45.77\pm0.49$ & $204.70\pm0.70$ & $77\pm1$ \\ 
H$\delta$  4101.735   & $-47.42\pm0.18$ & $205.81\pm0.25$ & $659\pm3$ \\ 
He\,{\sc i}  4143.759 & $-49.44\pm0.87$ & $198.33\pm1.18$ & $55\pm2$ \\ 
Ca\,{\sc i}  4226.728 & $-45.78\pm1.44$ & $209.03\pm1.95$ & $36\pm2$ \\ 
Fe\,{\sc i}  4232.726 & $-47.77\pm1.15$ & $201.03\pm1.55$ & $41\pm2$ \\ 
Fe\,{\sc i}  4266.964 & $-46.93\pm0.92$ & $198.87\pm1.26$ & $54\pm2$ \\ 
H$\gamma$ 4340.465    & $-47.48\pm0.16$ & $207.11\pm0.22$ & $842\pm3$ \\ 
Fe\,{\sc i}  4351.544 & $-48.56\pm1.04$ & $201.36\pm1.43$ & $63\pm2$ \\ 
He\,{\sc i}  4387.928 & $-46.72\pm0.61$ & $197.68\pm0.84$ & $84\pm2$ \\ 
He\,{\sc i}  4471.480 & $-46.99\pm0.32$ & $202.90\pm0.46$ & $144\pm2$ \\ 
Mg\,{\sc ii} 4481.130 & $-46.51\pm0.72$ & $201.59\pm1.04$ & $71\pm2$ \\ 
Fe\,{\sc i}  4549.467 & $-46.01\pm1.05$ & $201.43\pm1.43$ & $49\pm2$ \\ 
He\,{\sc i}  4713.146 & $-47.44\pm0.49$ & $204.45\pm0.69$ & $75\pm2$ \\ 
H$\beta$ 4861.327     & $-47.39\pm0.16$ & $208.67\pm0.23$ & $1406\pm4$ \\ 
He\,{\sc i}  4921.929 & $-48.30\pm0.30$ & $204.02\pm0.43$ & $149\pm2$ \\ 
Fe\,{\sc ii} 4923.921 & $-47.35\pm0.52$ & $204.94\pm0.73$ & $80\pm2$ \\ 
Fe\,{\sc i}  4957.597 & $-46.94\pm0.98$ & $205.54\pm1.44$ & $59\pm2$ \\ 
He\,{\sc i}  5015.675 & $-48.93\pm0.26$ & $205.02\pm0.37$ & $197\pm2$ \\ 
Fe\,{\sc ii} 5018.434 & $-48.60\pm0.50$ & $203.93\pm0.73$ & $105\pm2$ \\ 
Mg\,{\sc i}  5167.322 & $-49.24\pm0.63$ & $205.19\pm0.90$ & $90\pm2$ \\ 
Fe\,{\sc i}  5168.898 & $-45.00\pm0.51$ & $207.13\pm0.73$ & $112\pm2$ \\ 
Mg\,{\sc i}  5172.684 & $-46.67\pm0.96$ & $207.01\pm1.41$ & $127\pm2$ \\ 
Mg\,{\sc i}  5183.604 & $-48.44\pm0.66$ & $205.68\pm0.99$ & $136\pm3$ \\ 
Fe\,{\sc i}  5227.189 & $-46.44\pm0.76$ & $200.01\pm1.10$ & $70\pm2$ \\ 
Fe\,{\sc i}  5269.537 & $-48.88\pm0.85$ & $198.55\pm1.27$ & $92\pm2$ \\ 
Fe\,{\sc ii} 5276.002 & $-48.15\pm0.77$ & $202.46\pm1.11$ & $80\pm2$ \\ 
Fe\,{\sc ii} 5316.615 & $-46.43\pm0.57$ & $201.34\pm0.82$ & $102\pm2$ \\ 
Fe\,{\sc i}  5328.038 & $-47.10\pm0.67$ & $203.37\pm0.99$ & $126\pm2$ \\ 
Fe\,{\sc i}  5405.774 & $-48.88\pm1.95$ & $205.80\pm1.59$ & $51\pm2$ \\ 
Fe\,{\sc i}  5429.695 & $-46.80\pm0.96$ & $207.31\pm1.43$ & $73\pm2$ \\ 
Fe\,{\sc i}  5434.524 & $-48.12\pm2.77$ & $198.15\pm3.37$ & $25\pm2$ \\ 
Fe\,{\sc i}  5446.871 & $-45.26\pm0.97$ & $206.78\pm1.41$ & $53\pm2$ \\ 
Fe\,{\sc i}  5455.609 & $-46.39\pm0.82$ & $204.71\pm1.20$ & $62\pm2$ \\ 
He\,{\sc i}  5875.618 & $-47.07\pm0.29$ & $207.82\pm0.41$ & $623\pm7$ \\ 
Na\,{\sc i}  5889.950 & $-48.81\pm1.07$ & $211.58\pm1.49$ & $160\pm8$ \\ 
Na\,{\sc i}  5895.924 & $-49.29\pm1.13$ & $209.46\pm1.59$ & $124\pm8$ \\ 
Fe\,{\sc i}  6136.994 & $-45.82\pm3.19$ & $207.64\pm3.19$ & $112\pm8$ \\ 
Ca\,{\sc i}  6162.170 & $-49.04\pm2.86$ & $210.47\pm2.27$ & $31\pm5$ \\ 
Fe\,{\sc i}  6191.588 & $-48.29\pm1.77$ & $205.30\pm2.30$ & $47\pm7$ \\ 
Mg\,{\sc ii} 6346.962 & $-45.17\pm0.89$ & $203.44\pm1.21$ & $123\pm6$ \\ 
Fe\,{\sc i}  6393.601 & $-46.88\pm2.70$ & $198.16\pm3.22$ & $50\pm7$ \\ 
Fe\,{\sc i}  6400.001 & $-46.92\pm1.74$ & $208.27\pm2.57$ & $57\pm6$ \\ 
Fe\,{\sc ii} 6456.383 & $-46.43\pm1.70$ & $204.89\pm2.21$ & $125\pm7$ \\ 
Ca\,{\sc i}  6462.567 & $-47.03\pm1.90$ & $209.01\pm2.58$ & $37\pm5$ \\ 
Fe\,{\sc i}  6494.980 & $-46.55\pm1.30$ & $209.76\pm1.74$ & $56\pm5$ \\ 
\end{tabular}
\end{table*}

\begin{table*}
\centering
\begin{tabular}{@{}lccc@{}}
H$\alpha$ 6562.760    & $-46.77\pm0.25$ & $210.73\pm0.33$ & $4073\pm13$ \\ 
He\,{\sc i}  6678.149 & $-47.50\pm0.51$ & $206.26\pm0.61$ & $645\pm7$ \\ 
He\,{\sc i}  7065.188 & $-46.25\pm0.54$ & $205.69\pm0.64$ & $820\pm7$ \\ 
He\,{\sc i}  7281.349 & $-48.65\pm0.64$ & $202.81\pm0.82$ & $344\pm8$ \\ 
O\,{\sc i}   7771.944 & $-47.61\pm0.61$ & $203.64\pm0.81$ & $223\pm5$ \\ 
O\,{\sc i}   7774.166 & $-48.06\pm1.39$ & $202.84\pm1.12$ & $234\pm5$ \\ 
O\,{\sc i}   7775.388 & $-46.94\pm1.55$ & $204.89\pm1.38$ & $131\pm5$ \\ 
Mg\,{\sc ii} 7896.368 & $-48.23\pm0.90$ & $202.55\pm1.30$ & $131\pm7$ \\ 
Na\,{\sc i}  8183.256 & $-49.93\pm1.08$ & $206.13\pm1.62$ & $74\pm6$ \\ 
Na\,{\sc i}  8194.824 & $-46.10\pm1.46$ & $206.74\pm2.13$ & $89\pm7$ \\ 
Fe\,{\sc i}  8327.056 & $-47.02\pm1.04$ & $200.82\pm1.54$ & $158\pm8$ \\ 
C\,{\sc i}   8335.150 & $-46.96\pm0.69$ & $207.21\pm0.94$ & $283\pm7$ \\ 
O\,{\sc i}   8446.359 & $-47.51\pm0.73$ & $206.06\pm0.98$ & $340\pm9$ \\ 
Ca\,{\sc ii} 8498.020 & $-46.89\pm0.19$ & $207.18\pm0.27$ & $925\pm9$ \\ 
Ca\,{\sc ii} 8542.090 & $-47.39\pm0.15$ & $207.64\pm0.21$ & $1322\pm9$ \\ 
P14 8598.392          & $-48.88\pm2.00$ & $200.56\pm2.78$ & $256\pm12$ \\ 
Ca\,{\sc ii} 8662.140 & $-47.65\pm0.19$ & $207.95\pm0.29$ & $901\pm9$ \\ 
P13 8665.019          & $-45.51\pm1.88$ & $202.98\pm2.68$ & $489\pm12$ \\ 
Fe\,{\sc i}  8688.625 & $-48.01\pm1.07$ & $203.53\pm1.66$ & $176\pm9$ \\ 
P12 8750.473          & $-44.68\pm1.71$ & $203.42\pm2.50$ & $719\pm13$ \\ 
Mg\,{\sc i}  8806.757 & $-46.67\pm0.72$ & $205.38\pm0.99$ & $244\pm8$ \\ 
P11 8862.784          & $-45.69\pm1.35$ & $202.23\pm1.93$ & $1010\pm13$ \\ 
Ca\,{\sc ii} 8912.070 & $-49.99\pm0.78$ & $202.51\pm1.19$ & $194\pm9$ \\ 
Ca\,{\sc ii} 8927.360 & $-49.91\pm0.76$ & $202.22\pm1.10$ & $207\pm9$ \\ 
P10 9014.911          & $-47.08\pm0.66$ & $201.91\pm0.95$ & $1337\pm13$ \\ 
C\,{\sc i}   9061.430 & $-46.69\pm0.88$ & $202.20\pm1.12$ & $638\pm10$ \\ 
C\,{\sc i}   9078.280 & $-46.30\pm0.85$ & $203.85\pm1.08$ & $255\pm9$ \\ 
C\,{\sc i}   9088.510 & $-47.30\pm0.79$ & $208.63\pm1.07$ & $378\pm10$ \\ 
C\,{\sc i}   9094.830 & $-45.99\pm0.90$ & $201.89\pm1.19$ & $385\pm10$ \\ 
C\,{\sc i}   9111.800 & $-48.71\pm0.79$ & $206.42\pm1.01$ & $281\pm9$ \\ 
Fe\,{\sc i}  9212.981 & $-48.42\pm1.14$ & $207.64\pm1.67$ & $212\pm12$ \\ 
Mg\,{\sc ii} 9218.248 & $-46.67\pm0.48$ & $204.95\pm0.70$ & $493\pm12$ \\ 
P9  9229.015          & $-49.80\pm0.54$ & $203.76\pm0.78$ & $2079\pm20$ \\ 
Mg\,{\sc ii} 9244.266 & $-46.71\pm0.63$ & $203.53\pm0.91$ & $289\pm11$ \\ 
C\,{\sc i}   9405.730 & $-47.00\pm0.77$ & $208.83\pm1.01$ & $799\pm19$ \\ 
P$\epsilon$ 9545.972  & $-47.40\pm0.43$ & $205.84\pm0.62$ & $2825\pm26$ \\ 
C\,{\sc i}   9620.800 & $-48.95\pm1.00$ & $204.99\pm1.41$ & $243\pm15$ \\ 
C\,{\sc i}   9658.440 & $-45.49\pm0.99$ & $208.87\pm1.40$ & $202\pm18$ \\ 
P$\delta$ 10049.37    & $-48.12\pm0.72$ & $206.70\pm0.65$ & $3576\pm60$ \\
\hline
\end{tabular}
\end{table*}

\label{lastpage}

\end{document}